\documentstyle[12pt]{article}
\newcommand{\be}{\begin{equation}}
\newcommand{\ee}{\end{equation}}
\newcommand{\bea}{\begin{eqnarray}}
\newcommand{\eea}{\end{eqnarray}}
\newcommand{\sirc}[1]{\stackrel{\circ}{#1}}

\def\id{\protect{{1 \kern-.28em {\rm l}}}}

\makeatletter \@addtoreset{equation}{section} \makeatother

\def\appendix#1{
  \setcounter{section}{1}
  \setcounter{equation}{0}
  \renewcommand{\thesection}{\Alph{section}}
  \section*{Appendix \thesection\protect\indent \parbox[t]{11.715cm} {} }
  \addcontentsline{toc}{section}{Appendix \thesection\ \ \ }}

\begin{document}
\begin{flushright}
YITP-SB-00-02\\ hep-th/0002028\\
\end{flushright}

\begin{center}
\LARGE{On the nonlinear KK reductions on spheres of supergravity
theories} \vskip1truecm

{\large\bf Horatiu Nastase and
 Diana Vaman}
\footnote{Research supported by National Science Foundation Grant
Phy 9722101\protect\\ E-mail addresses:
hnastase@insti.physics.sunysb.edu,
dvaman@insti.physics.sunysb.edu\protect } \\ {\large C.N.Yang
Institute for Theoretical Physics,\\ SUNY Stony Brook, NY
11794-3840, USA\\ } \vskip1truecm
\end{center}
\abstract{
\parbox {4.75 in}{

~~~~~We address some issues related to the  construction of general 
Kaluza-Klein (KK) ans\"atze for the 
 compactification of a supergravity (sugra) theory on a sphere $S_m$.
We first reproduce various ans\"atze for compactification to 7d from the 
ansatz for the full nonlinear KK reduction of 11d sugra on $AdS_7\times S_4$.
As a side result, we obtain a lagrangean formulation of 7d ${\cal N}=2$
gauged sugra, which so far had only a on-shell formulation,
through field equations and constraints.
The $AdS_7\times S_4$ ansatz generalizes therefore all previous sphere 
compactifications to 7d.
Then we consider the case when the scalars in the lower dimensional
theory are in a coset $Sl(m+1)/SO(m+1)$, and
we keep the maximal gauge group $SO(m+1)$.
The 11-dimensional sugra truncated
on $S_4$ fits precisely the case under consideration, and serves as a model for
our construction. We find that the metric ansatz has a universal
expression, with the internal space deformed by the scalar fluctuations to a
conformally rescaled ellipsoid. 
We also find the ansatz for the dependence of the antisymmetric tensor
on the scalars. We  comment on
the fermionic ansatz, which will contain a matrix $U$ interpolating between 
the spinorial
$SO(m+1)$ indices of the spherical harmonics  and the $R$-symmetry indices 
of the fermionic fields in the lower dimensional sugra theory. We derive 
general conditions
which the matrix $U$ has to satisfy and we give a formula for the vielbein 
in terms of $U$. As an application of our methods we obtain the full ansatz 
for the metric and vielbein for 10d sugra on $AdS_5\times S_5$ (with no 
restriction on any fields).
}
\newpage

\section{Introduction}

The problem of Kaluza-Klein (KK) reduction of sugras on spheres 
is a very difficult one. 
In general, one obtains a gauged sugra after KK reduction, and finding the 
nonlinear ansatz which realizes the embedding of this gauged sugra into 
the original model is quite nontrivial. In the case of the reduction 
of 11d sugra on $AdS_4\times S_7$, parts of the ansatz where found in 
\cite{dwn84,dwnw,dwn86,dwn87}. 
In particular, the ansatz for the antisymmetric tensor was not explicitly
obtained (only in a highly implicit form, not suitable for calculations). 
 The nonlinear ansatz for the $AdS_7\times S_4$ reduction of 
11d sugra was found by the present authors
 together with Peter van Nieuwenhuzen in \cite{nvv,nvv2}, and was 
the first to give a complete solution for the nonlinear KK ansatz for a
sphere compactification. The ansatz for the compactification of 10d IIB 
sugra on $AdS_5\times S_5$ is the most notable absence in this respect.
Another problem which has to be addressed is the {\em consistency} of the 
truncation (KK reduction). In \cite{dwn87}, 
the consistency of the KK reduction of 
an SU(8)-invariant version of the 11d sugra was given. In \cite{nvv2}, 
we gave a 
proof of consistency of the reduction of the usual 11d sugra on $AdS_7\times
S_4$. In this paper, we will not have anything to say about the 
consistency of the ans\"atze we will propose, except when we take various 
further truncations of our $AdS_7\times S_4$ ansatz to compare to other work in
the literature.

The recent interest in sphere compactifications came from the AdS-CFT
correspondence \cite{mald,gkp,witten,3point,4point,asmo}. 
A lot of work has been done in finding ansatze for 
further truncations of the models after the KK reduction on spheres. 
In \cite{cdh,cllp} it was considered the further truncation of the KK 
reduction of 11d and 10d IIB sugras on $AdS_4\times S_7, AdS_7\times S_4$
and $AdS_5\times S_5$ to subsets of U(1) gauge fields and scalars. In 
\cite{lp} was studied 
the truncation of the KK reduction of 11d sugra on $AdS_7\times 
S_4$ to the bosonic sector of an N=2 (SU(2)) gauged sugra. A set of only 
gravity and scalar fields was considered in \cite{cglp}, whereas \cite{cjlp}
analyzed various compactifications giving rise to gravity plus scalar fields
in a coset manifold.
In \cite{clp} it was found an explicit ansatz for the embedding of the 
N=4 (SO(4)) gauged sugra in 11d (via $S_7$ compactification).
 In \cite{volk,chavo} the KK reduction of 10d sugra 
on $S_3\times S_3$ was analyzed, and in \cite{lpt} the embedding of N=4 
gauged sugra into 10d sugra.  

In this paper we want to generalize some of the results obtained in 
\cite{nvv2}, most 
notably for application to the $AdS_4\times S_7$ and $AdS_5\times S_5$ cases, 
and also to relate our results to the other works  in the literature.
We will explicitly show that the other ans\"atze for embedding of subsets 
of fields of the 7d N=4 gauged sugra into 11 dimensions can be obtained as 
particular cases of the ansatz in \cite{nvv2}. More precisely, we will 
recover: the ansatz involving 
the graviton and 4 scalars in \cite{cglp}, the ansatz involving the graviton,
an SU(2) gauge field, a scalar and an antisymmetric tensor in \cite{lp},
the ansatz for the $S_3$ compactification of 10 d sugra of \cite{cs} 
(which leads to $d=7$ ${\cal N}=2$ gauged sugra without topological mass 
term) 
and the ansatz involving a graviton, two abelian gauge fields and two scalars 
in \cite{cdh}. The correspondence between 
our ansatz truncated to the fields of $d=7$ ${\cal N}=2$ gauged sugra 
\cite{tn83} and the 
ansatz of L\"u and Pope \cite{lp} is discussed in detail. To obtain agreement 
between the two ansatze (and between the field equations derived from
the truncated action of the ${\cal N}=4$ model and the field equations of 
${\cal N}=2$ gauged sugra), we need to make a 
redefinition of the three index potential $A_{(3)}$ (in the ${\cal N}=2$ model)
as a linear combination of the three index potential $S_{(3)}$ (in the ${\cal 
N}=4$ model) and the Chern Simons form $\omega_{(3)}$.  
We noted that there is a lagrangean formulation of $d=7$ ${\cal N}=2$ gauged 
sugra, obtained from the maximal 7d gauged sugra lagrangean by truncatation,
which yields the set of field equations and constraints which so far had been
used to describe the ${\cal N}=2$ model.
By taking a singular limit of the ansatz which yields ${\cal N}=2$ $d=7$ 
gauged sugra from $d=11$ sugra with topological mass term \cite{lp, nvv2}, 
we obtain an 
ansatz which yields ${\cal N}=2$ $d=7$ gauged sugra without topological mass 
term. Thus we confirm the consistency of the ansatz proposed by 
Chamseddine and Sabra \cite{cs} for the $S_3$ compactification of type
I 10d sugra to 7d, who derived their ansatz only from the 
requirement to reproduce the action of gauged ${\cal N}=2$ $ d=7$ sugra 
(which does not necessarily imply consistency). Since we obtain it as a 
singular limit from a consistent ansatz we implicitly prove the 
consistency of the truncation in \cite{cs}.

The generalizations we are interested in are related to the ansatz for the 
embedding of the lower (d-) dimensional scalar fields into the higher dimension
(D). We will see that when one compactifies on a sphere $S_n$, 
one has a scalar coset which is at least $Sl(n+1)/SO(n+1)$. We will mostly 
work with this coset, but for the $AdS_4\times S_7$ and $AdS_5\times S_5$ 
compactifications we will analyze also the case of the full coset. 
One would think that only the cases (d,n)= (4,7),(7,4) and (5,5) are of 
interest, and the metric and antisymmetric tensor ansatz for this case
were found in \cite{cglp}, but actually the case of general d and n has 
its own  interest. That is for instance because it was found useful 
in problems related to the Randall-Sundrum model \cite{rs1,rs2}, for instance
see \cite{bbs}. For Randall-Sundrum type scenarios, one needs models of gravity
interacting with scalar fields. Since finding a consistent realization of 
this scenario inside a susy theory is an (yet) unsolved problem (see
\cite{wz,kl,bc}), it is worth studying general gravity + scalar systems. 
Moreover, \cite{cglp} considered  only n scalars, which is OK
as long as no gauge fields couple to them. But we find the metric and 
antisymmetric tensor ansatz and show how to couple the gauge fields to 
the scalars.  A completely new result is that we obtain the full metric ansatz
for the KK reduction of 10d IIB sugra on $AdS_5\times S_5$, which means that 
if one has a solution of 5d gauged sugra one can find the metric of the 
full 10d solution, and sometimes that is all one needs (for instance for the 
applications for the Randall-Sundrum type scenarios). 

Yet another extension of the ansatz in \cite{nvv2} is related to the fermions.
We see in general that there will be a scalar field-dependent
matrix $U$ which relates the fermions to the Killing spinors, as was done 
in \cite{dwn84} for the $AdS_4\times S_7$ case and in \cite{nvv,nvv2} for 
the $AdS_7\times S_4$ case (the matrix U in the 4d case was 
introduced also in \cite{anp,nilsson}). We find an ansatz 
for the vielbein which depends on the matrix $U$ and an equation which $U$ 
has to satisfy. We analyze separately the $AdS_4\times S_7$ and $AdS_5\times 
S_5$ cases, and we find there the vielbein and the U equation for the theory 
with the full scalar coset (no restriction to $Sl(n+1)/SO(n+1)$).

The paper is organized as follows. In section 2 we derive from the ansatz
\cite{nvv2} for the consistent KK reduction of $d=11$ sugra to maximal $d=7$ 
gauged sugra the other ans\"atze 
in the literature: \cite{lp,cglp,cdh,cs,cllp}}.
for the KK reduction to smaller bosonic sectors of the maximal gauged sugra in
$d=7$. In section 3 we write the metric ansatz for the $S_n$ 
compactification involving a scalar coset $Sl(n+1)/SO(n+1)$, and the 
dependence of the antisymmetric tensor field on the scalars. For the 
$AdS_4\times S_7$ and the $AdS_5\times S_5$ cases, we get the full metric 
ansatz (with no restrictions). In section 4 we say a few words about the 
fermionic ansatz, introduce the matrix $U$, and derive the vielbein and 
$U$ equation. Again, in the $AdS_4\times S_7$ and $AdS_5\times S_5$ cases,
we impose no restriction. We finish with discussions and conclusions. 
In the Appendix we 
give some useful things about the Killing spinors on spheres.


\section{From d=11 sugra to d=7 sugra \\~~~~~~~via $S_4$ compactification}

The consistent embedding of seven dimensional supergravity
theories in higher dimensions was thoroughly studied in the recent
literature. We will first describe the case of $S_4$ truncation of the
original d=11 supergravity \cite{nvv,nvv2}, and then obtain other embeddings
as particular cases. When compactifying on $S_4$ we obtain ${\cal N}=4$
maximal gauged sugra in seven dimensions \cite{ppn}. The scalars
${\Pi_A}^i(y)$ parameterize an $Sl(5,\bf {R})/SO(5)_c$ coset
manifold. The gauge fields $B^{AB}_\alpha(y)$ are in the adjoint
of $SO(5)_g$, and the gravitini and spin 1/2 fields transform in
the spinorial representation of the R-symmetry group $SO(5)_c$ and
in the vector-spinor representation, respectively. As we shall
see, this model admits consistent truncations to smaller field
subsets: ${\cal N}=2$ gauged sugra, \footnote{We are thankful to
C.N.Pope H.L\"u and A.Sadrzadeh for pointing to us this fact.} or
purely bosonic sectors such as the graviton and the whole set of
scalars, or the graviton, two abelian gauged fields (in the Cartan
subalgebra of $SO(5)$) and two scalars. Our ansatz for the metric
factorizes into a rescaled 7 dimensional metric and a gauge
invariant two form which depends on the scalar fields through the
composite tensor $T^{AB}(y)={{\Pi^{-1}}_i}^A{{\Pi^{-1}}_i}^B$.
\bea 
ds_{11}^2(x,y)&=&\Delta^{-2/5}(x,y)\left[ g_{\alpha\beta}(y)
dy^{\alpha}dy^{\beta}+\frac{1}{m^2}(DY)^A \frac{T^{-1}_{AB}(y)}{Y\cdot
T(y)\cdot Y} (DY)^B\right]\label{metric}
\\
DY^A&=&dY^A+2mB^{AB}(y) Y_B\nonumber\\ 
\eea 
The scale factor is
\be
\Delta^{-6/5}(x,y)=Y^A(x)T_{AB}(y)Y^B(x) 
\ee 
The full  internal metric $g_{\mu\nu}= m^{-2}\Delta^{4/5}\partial _{\mu}Y^A
T^{-1}_{AB}\partial_{\nu}Y^B$ describes a conformally rescaled
ellipsoid: $g_{\mu\nu}=m^{-2}\Delta^{4/5}\partial_{\mu}Z^A
\partial_{\nu}Z_A$ where the $Z^A(x)$'s are constrained to lie on the
ellipsoid. Therefore the overall effect of all scalar fluctuations
on the internal metric is to deform the background sphere $\sirc
g_{\mu\nu}=m^{-2}
\partial_\mu Y^A \partial_\nu Y^B \delta_{AB}$, $Y^A(x) Y^B(x)\delta_{AB}=1$
into a conformally rescaled ellipsoid.  Other types of fluctuations would 
correspond to 'massive' modes.

The 4-form field strength of the three index ``photon'' of the standard 
$d=11$ sugra is given by 
\bea
\frac{\sqrt{2}}{3}F_{(4),11}&=&\epsilon_{A_1...A_5}\left[-\frac{1}{3m^3}
(DY)^{A_1} ...(DY)^{A_4}\frac{(T\cdot Y)^{A_5}}{Y\cdot T\cdot Y}
\right.\nonumber\\&&\left. +\frac{4}{3m^3}
(DY)^{A_1}(DY)^{A_2}(DY)^{A_3}D\left(\frac{(T\cdot Y)^{A_4}}{
Y\cdot T\cdot Y}\right)Y^{A_5}\right.\nonumber\\
&&\left.+\frac{2}{m^2}F_{(2)}^{A_1A_2}(DY)^{A_3}(DY)^{A_4}\frac{(T\cdot
Y)^{A_5}}{Y\cdot T\cdot Y}+\frac{1}{m}
F_{(2)}^{A_1A_2}F_{(2)}^{A_3A_4}Y^{A_5}\right]\nonumber\\&&
+d(S_{(3)B}Y^B)
\label{f} 
\eea 
where we define the Yang-Mills field strength to be $F_{(2)}^{AB}=2(dB_{(1)}
+(B_{(1)}\cdot B_{(1)})^{AB}$ and the independent fluctuation form 
$S_{(3)B\alpha\beta\gamma}=-\frac{8i}{\sqrt{3}}S_{\alpha\beta\gamma ,B}$ 
is real. (We also used the implicit convention $F_{(4),11}=F_{\Lambda\Pi
\Sigma\Omega} dx^\Lambda\wedge dx^\Pi\wedge dx^\Sigma \wedge dx^\Omega$.)

Finally, in order to produce a first order field equation for the 
antisymmetric tensor field $S_{\alpha\beta\gamma}$ (self-duality in odd 
dimensions \cite{nt}) as well as for the 
consistency of the fermionic susy transformation laws we had to
introduce in the $d=11$ sugra model an auxiliary field ${\cal B}_{MNPQ}$.
The ansatz for the KK truncation of the 4-index auxiliary field has a single
nonvanishing component
\be
\frac{{\cal B}_{\alpha\beta\gamma\delta}}{\sqrt{g_{11}}}=
\frac{i}{2\sqrt 3}\epsilon_{\alpha\beta\gamma\delta\epsilon\eta\zeta}\frac{
\delta S_7}{\delta S_{\epsilon\eta\zeta , A}} Y^A
\ee
where $S_7$ is the action of the 7-dimensional gauged sugra.
The other models studied in the literature (in $d=7$) are special
cases of this general ansatz.
\begin{itemize}
\item{graviton and scalars in $Sl(5)/SO(5)$}\cite{cglp}
\bea ds_{11}^2&=&
\tilde\Delta^{1/3}ds_7^2+\frac{1}{g^2}\tilde\Delta^{-2/3}\sum_i
X_i^{-1} d\mu_i^2\label{metric1}\\
F_{(7),11}&=&g\sum_{i}(2X_i^2\mu_i^2-\tilde\Delta X_i)
\epsilon_{(7)}- \frac{1}{2g}\sum_i X_i^{-1}\star dX_i\wedge
d(\mu_i^2)\label{f2} 
\eea 
where the scalar fields satisfy the
constraint $\prod_{i=1}^5 X_i =1$, and spherical harmonics $\mu_i$
are such that $\sum_{i=1}^5 \mu_i^2=1$. The conformal factor
$\tilde\Delta$ is defined as $\tilde\Delta=\sum_{i=1}^5 X_i
\mu_i^2$. We easily recognize that the two sets of spherical
harmonics $\mu_i$ and our $Y^A$ are the same. Moreover, the scalar
fields $X_i$ are embedded in $T_{AB}$ as
\be
T_{AB}=diag(X_1, X_2,\dots,X_5)\label{Tx} 
\ee 
Therefore, $\Delta^{-6/5}=\tilde\Delta$, and the metric ansatz given in
(\ref{metric1}) coincides with the ansatz (\ref{metric}) if we
make the identification (\ref{Tx}), and we set the gauge fields to
zero. However, it is not immediately obvious that the ansatz for
the antisymmetric tensor in (\ref{f2}) also coincides with
(\ref{f}), upon truncation. To check agreement between the two
ansatze we compute the dual of $F_{(7),11}$: 
\bea 
F_{(4),11}&=&\hat\star
F_{(7),11}=\nonumber\\
&=& \frac{\sqrt{g_{11}}}{4!}{\epsilon^{\alpha_1 \dots
\alpha_7}}_{\mu\nu\rho\sigma}g\sum_i (2X_i^2\mu_i^2-\tilde\Delta
X_i) \frac{\sqrt{g_{7}}}{7!}\epsilon_{\alpha_1 \dots\alpha_7}
dx^\mu dx^\nu dx^\rho dx^\sigma\nonumber\\
&&-\frac{\sqrt{g_{11}}}{4!} {\epsilon^{\alpha_1 \dots
\alpha_6\mu}}_{\alpha_7\nu\rho\sigma}\sum_i X_i^{-1}
\partial_\alpha X_i
\partial_\mu(\mu_i^2)\frac{\sqrt{g_7}}{6!2 g}{\epsilon_{\alpha_1\dots
\alpha_6}}^\alpha
dy^{\alpha_7}dx^\nu dx^\rho dx^\sigma\nonumber\\ 
\eea 
where the notation $\hat\star$ means Hodge dual in the higher dimensional
space.

Using that the inverse metric is
\bea
g_{11}^{\alpha\beta}&=&g_7^{\alpha\beta} \tilde\Delta^{-1/3}\nonumber\\
g_{11}^{\mu\nu}&=&g^2\tilde\Delta^{-1/3}\left[\frac{1}{4}
\sum_i(\sirc\partial^\mu (\mu_i^2)
X_i)\sum_j (\sirc\partial^\nu (\mu_j^2)
X_j)-\tilde\Delta(\sum_i\sirc\partial^\mu (\mu_i) X_i
\sirc\partial^\nu (\mu_i))\right]
\nonumber\\{}\eea 
where $\sirc\partial^\mu = \sirc
g^{\mu\nu} \partial_\nu$ and $\sirc g^{\mu\nu}$ is the inverse
background metric on $S_4$, and  that
$\sqrt{g_{11}}=\sqrt{g_4}\sqrt{g_7} \tilde
\Delta^{7/6}=\tilde\Delta^{1/3} \sqrt{g_7}\sqrt{\sirc g_4}g^{-4}$ 
we get 
\bea
F_{(4)}&=&\epsilon_{\mu\nu\rho\sigma}\frac{\sqrt{\sirc
g_4}}{4!}\tilde \Delta^{-2}g^{-3} \sum_i(2X_i^2\mu_i^2-\tilde\Delta
X_i)dx^\mu dx^\nu dx^\rho dx^\sigma\nonumber\\
&&+\epsilon_{\mu'\nu\rho\sigma}\frac{\sqrt{\sirc g_4}}{4!}
\frac{\tilde\Delta^{-2} \tilde\Delta^{1/3}}{2g^3}
g^{\mu\mu'}\sum_i X_i^{-1}\partial_\alpha X_i
\partial_\mu(\mu_i^2)dy^{\alpha}dx^\nu dx^\rho dx^\sigma\label{f22}
\eea 
Using further that $\partial_\mu\mu_i\partial_\nu\mu_i=\sirc
g_{\mu\nu}$ and that \cite{nvv2}
\be
\epsilon_{ABCDE} dY^A \dots dY^D=\sqrt{\sirc g_4} \epsilon_{\mu\nu
\rho\sigma} Y^E dx^\nu dx^\rho dx^\sigma 
\ee 
we can show that (\ref{f22}) is a particular case of (\ref{f}).
\item{graviton, SU(2) gauge fields $A_{(1)}^i$, a three form $A_{(3),7}$ 
and one scalar $X$}\\
The ansatz for the consistent bosonic truncation of $d=11$ sugra
to ${\cal N} =2$ $d=7$ gauged sugra \cite{tn83} was given by H.L\"u and
C.N.Pope \cite{lp} 
\bea 
ds_{11}^2&=&\tilde\Delta^{1/3}ds_7^2
+\frac{2}{g^2} X^3\tilde\Delta ^{1/3}
d\xi^2+\frac{1}{2g^2}\tilde\Delta^{-2/3}X^{-1} \cos^2\xi\sum_i
(\sigma^i- gA_{(1)}^i)^2 \nonumber\\{}\label{spl}\\
F_{(4),11}&=&-\frac{1}{2\sqrt 2g^3}(X^{-8}
\sin^2\xi-2X^2\cos^2\xi+3X^{-3} \cos^2
\xi-4X^{-3})\nonumber\\
&&\tilde\Delta^{-2}\cos^3\xi
d\xi\wedge\epsilon_{(3)}-\frac{5}{2\sqrt
2g^3}\tilde\Delta^{-2}X^{-4}\sin\xi\cos^4\xi dX\wedge
\epsilon_{(3)} +\sin\xi F_{(4),7}\nonumber\\ &+&\frac{\sqrt
2}{g}\cos\xi X^4\star F_{(4),7}\wedge d\xi-\frac{1}{\sqrt 2g^2}
\cos\xi F_{(2)}^i \wedge d\xi\wedge h^i\nonumber\\
&-&\frac{1}{4\sqrt 2g^2}X^{-4}\tilde\Delta^{-1}\sin\xi\cos^2\xi
F_{(2)}^i \wedge h^k\wedge h^k\epsilon_{ijk}\label{f1} 
\eea 
where the 4-sphere is described as a foliation of 3-spheres,
parametrized by the Euler angles $(\theta,\varphi,\psi)$ with
latitude coordinate $\xi$; $\sigma^i$ are left invariant one forms
on $S_3$: $d\sirc s_4^2=d\xi^2+ 1/4\cos^2\xi \sigma^i \sigma^i$;
$F_{(2)}^i=dA_{(1)}^i +g/2\epsilon_{ijk}A_{(1)}^j A_{(1)}^k$ are
the Yang Mills field strengths with $g$ the coupling constant 
and $h^i=\sigma^i-gA_{(1)}^i$ are
the gauge invariant one forms. $\omega_{(3)}$ is a Chern-Simons
form, $d\omega_{(3)}=F_{(2)}^i\wedge F_{(2)}^i$. As discussed in
\cite{nvv2} the maximal gauge field group inherited from the
4-sphere, $SO(5)$, needs to be broken first to $SO(4)$ by first
setting $B^{A5}=0$ and further, from $SO(4)$ to one of its $SU(2)$
subgroups, by imposing an antiself-duality condition. Thus the
embedding of the gauge fields is
\be
B^{A5}=0\;\;\; ;\;\;\;B^{\hat\mu\hat\nu}=-\frac{1}{2}
\epsilon^{\hat\mu\hat\nu\hat\rho
\hat\sigma}B_{\hat\rho\hat\sigma},
\hat\mu=1,4\;\;\; ;\;\;\;B^{i4}=-\frac{1}{2\sqrt{2}}A_{(1)}^i 
\label{bembed}
\ee 
The scalar field of ${\cal N}=2$ gauged sugra is embedded in the 
$Sl(5)/SO(5)$ coset of maximal ${\cal N}=4$ sugra as
\be
T_{AB}=diag(X,X,X,X,X^{-4}) 
\ee 
The three index form of maximal $d=7$ sugra $S_{(3)}^A$ is also 
truncated to a singlet under $SU(2)$
\be
S_{(3)}\equiv S_{(3)}^5\;\;\; ;\;\;\;0=S_{(3)}^{\hat\mu}\;,\; \hat\mu=1,4 
\ee
The correspondence between the Cartesian spherical harmonics $Y^A$
and the new parametrization of the 4-sphere is
\be
Y^5=\sin\xi\;\;\; ;\;\;\;Y^{\hat\mu}=\cos\xi \hat
Y^{\hat\mu}\;,\;\hat\mu=1,4 \ee where $\hat Y^{\hat\mu}$ are
constrained to lie on a 3-sphere. Finally, the Cartesian
coordinates $\hat Y$ are related to the Euler angles
$(\theta,\varphi,\psi)$ by 
\bea 
\hat Y^4+i\hat
Y^3=\cos{\frac\theta 2}\exp(\frac{i(\varphi+\psi)}{2})\nonumber
\\
\hat Y^2+i\hat Y^1=\sin{\frac \theta
2}\exp(\frac{i(\psi-\varphi)}{2}) 
\eea 
and by organizing them into an $SU(2)$ matrix
\be
{\cal G}=\pmatrix{\hat Y^4+i\hat Y^3& \hat Y^2+i\hat Y^1\cr 
-\hat Y^2+i\hat Y^1&\hat Y^4-i\hat Y^3 } =\hat Y^4\id +i\tau^k\hat Y^k
\ee 
we get the invariant
$SU(2)$ one forms $\sigma^i$ in terms of the Cartesian coordinates
from
\be
\sigma^i=\frac{1}{i} tr(\tau^i {\cal G}^{-1}d{\cal G})
=2(\hat Y^4d\hat Y^i-\hat Y^id\hat Y^4+\epsilon_{ijk}\hat Y^jd\hat Y^k)
\label{sigma} 
\ee 
Also, from (\ref{sigma}) we get
\be
d\sigma^{i}=4(-d\hat Y^i d\hat Y^4 +\frac{1}{2}\epsilon_{ijk}
d\hat Y^j d\hat Y^k) \label{dsigma}
\ee 
Thus, the equivalence between the two
metric ansatze is straightforward \cite{nvv2}. In (\ref{metric})
we substitute the truncated fields and the redefined spherical
harmonics and we get
\bea 
ds_{11}^2&=&\Delta^{-2/5}(x,y)\left[
g_{\alpha\beta}(y) dy^{\alpha}dy^{\beta}+\frac{1}{m^2}(DY)^A
\frac{T^{-1}_{AB}(y)}{Y\cdot T(y)\cdot Y} (DY)^B\right]\nonumber\\
&=&\tilde\Delta^{1/3} ds_7^2+m^{-2}\tilde\Delta^{1/3}X^3 d\xi^2+m^{-2}
\tilde\Delta^{-2/3}\cos^2\xi X^{-1} \left(d\hat Y^{\hat\mu}d\hat
Y^{\hat\mu} \right.\nonumber\\&+&\left. 2m(B\cdot \hat Y)^{\hat\mu} d\hat
Y^{\hat\mu} +m^2 (B\cdot \hat Y)^{\hat\mu} (B\cdot
\hat Y)^{\hat\mu} \right) 
\eea 
Using the relation (\ref{sigma}) and the anti-selfduality condition 
on the gauge fields, the 11-dimensional invariant line element becomes: 
\be
ds_{11}^2=\tilde\Delta^{1/3} ds_7^2+\frac{2}{g^2}\tilde\Delta^{1/3}X^3 d\xi^2+
\frac{1}{2g^2}\cos^2\xi X^{-1} (\sigma^i -gA_{(1)}^i)(\sigma^i
-gA_{(1)}^i) 
\ee 
and coincides with (\ref{spl}), provided that the 
Yang-Mills constant is $g=\sqrt{2}m$.

The ansatz for the 11 dimensional 3-index tensor
$A_{\Pi\Sigma\Omega}$ given in \cite{nvv2} differs from the one of
L\"u and Pope. However, since it is defined only up to a gauge
transformation (namely up to $\partial_{[\Pi}\Lambda_{\Sigma\Omega
]}$) we will compare the ansatz for its field strength
$F_{(4),11}=dA_{(4),11}$, i.e. (\ref{f}) vs. (\ref{f1}). The two
ans\"atze will turn out to be the same if we assume that the three form
$A_{(3)}$ is a linear combination of $S_{(3)}$ and 
$\omega_{(3)}$. 

To show the equivalence of the two ans\"atze, we
need to make use of the following identities involving the
spherical harmonics: 
\bea
\epsilon_{\hat\mu\hat\nu\hat\rho\hat\sigma}D\hat Y^{\hat\mu}\wedge
D\hat Y^{\hat\nu}\wedge D\hat Y^{\hat\rho} \wedge D\hat Y^{\hat\sigma}&=&0\\
\epsilon_{\hat\mu\hat\nu\hat\rho\hat\sigma}D\hat Y^{\hat\mu}\wedge
D\hat Y^{\hat\nu}\wedge D\hat Y^{\hat\rho} \hat Y^{\hat\sigma}&=&
-\frac{1}{8} \epsilon_{ijk} h^i \wedge h^j \wedge h^k 
\equiv-\frac{3!}{8}\epsilon_{(3)}\\
\epsilon_{\hat\mu\hat\nu\hat\rho\hat\sigma}D\hat Y^{\hat\mu}
\wedge D\hat Y^{\hat\nu}\wedge F_{(2)}^{\hat\rho \hat\sigma}
&=&-\frac{1}{2\sqrt{2}}\epsilon_{ijk} F_{(2)}^i \wedge h^j \wedge h^k\\
\epsilon_{\hat\mu\hat\nu\hat\rho\hat\sigma}F_{(2)}^{\hat\mu\hat\nu}
\wedge D\hat Y^{\hat\rho}  Y^{\hat\sigma}&=&\frac{1}{\sqrt{2}}F_{(2)}^i h^i
\eea
To prove these identities, one needs to use
the form of $\sigma ^i$ and $d\sigma ^i$ in (\ref{sigma},\ref{dsigma}) and
the embedding of the gauge fields in (\ref{bembed}) (remember that
$F_{(2)}^{\hat\mu\hat\nu}=2(dB_{(1)}^{\hat\mu\hat\nu}+2 mB_{(1)}^{\hat\mu
\hat\rho}\wedge B_{(1)}^{\hat\rho\hat\nu})$). 
Then, one finds the various terms in (\ref{f}) in terms of the left-hand 
side of these identities:
\bea
&&\epsilon_{A_1...A_5}DY^{A_1}\wedge...\wedge DY^{A_4}\frac{T\cdot Y^{A_5}}
{Y\cdot T\cdot Y}=-\frac{3!}{2}\cos^3\xi d\xi\wedge \epsilon_{(3)}\\
&&\epsilon_{A_1...A_5}DY^{A_1}\wedge DY^{A_2}\wedge DY^{A_3} \left(
\frac{T\cdot Y^{A_4}}{Y\cdot T\cdot Y}\right)Y^{A_5}=\frac{3!}{2}\cos^3\xi
\epsilon_{(3)}\wedge d\xi \tilde\Delta^{-2}\cos^3\xi \nonumber\\
&&\left[3X\tilde\Delta -2\sin^2\xi \cos^2\xi (X^{-8}-2 X^{-3} +X^2)+
X^2 \cos^2\xi \right.\nonumber\\
&&\left.+X^{-3} (\sin^4\xi \cos^3\xi +\cos^7\xi) +X^{-8}\sin^2\xi
\cos^5\xi   \right]\\
&&\epsilon_{A_1...A_5}F^{A_1A_2}_{(2)}\wedge DY^{A_3}\wedge DY^{A_4}
\frac{T\cdot Y^{A_5}}
{Y\cdot T\cdot Y}\nonumber\\
&&=2\epsilon_{\hat\mu\hat\nu\hat\rho\hat\sigma}\left(
\frac{\cos^2\xi \sin\xi X^{-4}}{\tilde{\Delta}}
F_{(2)}^{\hat\mu\hat\nu}\wedge
D\hat Y^{\hat\rho}\wedge D\hat Y^{\hat\sigma}
+2 \cos\xi d\xi\wedge F_{(2)}^{\hat\mu
\hat\nu}\wedge D\hat Y^{\hat\rho}\hat Y^{\hat\sigma}\right)
\nonumber\\
\eea
and one recovers most of (\ref{f1}).
\footnote{
To get the same overall coeficient as in (\ref{f1}) we have to take into 
account that the three form potential of 11-d sugra used in \cite{lp}
is related to the one used in \cite{cj, nvv} by a rescaling with $6\sqrt 2$,
and that in the conventions of \cite{lp} $F_{\Lambda\Pi\Sigma\Omega}$ equals
$4\partial_{[\Lambda}A_{\Pi\Sigma\Omega ]}$, while for the authors of \cite{cj,
nvv} $F_{\Lambda\Pi\Sigma\Omega}=24\partial_{[\Lambda}A_{\Pi\Sigma\Omega ]}$.
Also note that in \cite{nvv} the form $F_{(4)}$ was not normalized:
$F_{(4)}=F_{\Lambda\Pi\Sigma\Omega}dx^\Lambda \wedge dx^\Pi\wedge dx^\Sigma
\wedge dx^\Omega$.}

In the ${\cal N}=2$ model of \cite{tn83}
the lagrangean is quadratic in $dA_{(3)}$, and thus to reduce the on-shell
number of degrees of freedom of $A_{(3)}$ one needs to supplement the field
equations with 
the self-duality in odd dimensions constraint \cite{nt,lp}
\be
X^4\star F_{(4),7}=\frac{-1}{\sqrt 2}g A_{(3),7}+\frac{1}{2}\omega_{(3)}
\label{cons}
\ee
In fact, we should stress that the ${\cal N}=4$ model yields a first
order field equation for $S_{(3)}$, which is the square root of the quadratic
field equation of $A_{(3),7}$. We note that the field equations 
obtained from the truncated ${\cal N}=4$ action correspond to the field 
equations and constraints of the ${\cal N}=2$ model. This implies that
one should use the following lagrangean for describing the bosonic sector
of ${\cal N}=2$ $d=7$ gauged sugra with topological mass term
\footnote{Note that the limit $g\rightarrow 0$ is singular, as it was in
the ${\cal N}=4$ gauged sugra model.}:
\bea
\sqrt{g_7^{-1}}L_{7d, {\cal N}=2}&=&R+g^2(2X^{-3}+2X^2-X^{-8})\nonumber\\
&&+5\partial_
{\alpha}X^{-1}\partial_{\alpha}X+\frac{g^2}{2} X^{-4}S_{\alpha\beta\gamma}
S^{\alpha\beta\gamma}\nonumber\\
&&-\frac{1}{4}X^{-2}F_{\alpha\beta}^i F^{\alpha\beta \;i}
+\epsilon^{\alpha\beta\gamma\delta\epsilon\eta\zeta} \sqrt{g_7^{-1}}
\left(\frac{g}{24\sqrt 2}S_{\alpha\beta\gamma}F_{\delta\epsilon\eta\zeta}
\right.\nonumber\\&&+\left.
(-\frac{1}{48\sqrt 2 g}\omega_{\alpha\beta\gamma} +\frac{i}{8\sqrt{3}}
S_{\alpha\beta\gamma})F_{\delta\epsilon}^i F_{\eta\zeta}^i\right)	
\eea
The field equation for $S_{\alpha\beta\gamma}$ reads
\be
-g^2 X^{\-4} S_{(3)}=\star d(\sqrt{2}gS_{(3)}+\frac{i}{2\sqrt{3}}\omega_{3})
\ee
and coincides with the self-duality constraint (\ref{cons}) if 
\be
S_{(3)}=\frac{i}{2\sqrt 3}\left(A_{(3),7}-\frac{\omega_{(3)}}{\sqrt 2 g}\right)
\label{comb}
\ee
On the other hand, in $F_{(4)}$ (\ref{f}), the remaining terms which at 
the first sight seem to differ from the ones in (\ref{f1}) are 
\bea
&&\frac{3}{\sqrt 2}\left(\frac{-8i}{\sqrt{3}} d(S_{(3)}\sin\xi)
+\sqrt{2}{g}\sin\xi F_{(2)}^i\wedge F_{(2)}^i\right)\nonumber\\
&&=\frac{3}{\sqrt 2}\left(\frac{-8i}{\sqrt 6} \sin\xi 
d(\sqrt 2 S_{(3)}+\frac{i}{2\sqrt 3} 
\omega_{(3)})+\frac{8i}{\sqrt 3} S_{(3)} \cos\xi\wedge d\xi\right)\label{unu}\\
&&=\frac{3}{\sqrt 2}\left(\frac{8}{6} dA_{(3),7} \sin\xi +
\frac{8}{3g\sqrt{2}}\cos\xi X^4\star F_{(4),7}\wedge d\xi\right)\label{doi}
\eea
where to go from (\ref{unu}) to (\ref{doi}) we made use of the self-duality 
constraint (\ref{cons}) which is part of the truncation procedure for 
\cite{lp}. Thus we were able to completely recover the ansatz (\ref{f1})
from our ansatz (\ref{f}). This concludes that indeed, there exists a 
consistent truncation of the maximal gauged 7d sugra to ${\cal N}=2$ 
gauged ($SU(2)$) 7d sugra with topological mass term.


Finally, by taking the singular limit  $S_4\rightarrow
S_3\times {\bf R}$ in the same way as in \cite{cllp}, 
we recover the ansatz of Chamseddine and Sabra
\cite{cs}. They proposed an ansatz for obtaining the ${\cal N}=2$,
$d=7$ gauged sugra, with no topological mass term ($m=0$) for the
three form potential $A_{(3),7}$, and checked that the action of
lower dimensional theory is reproduced from the 11-dimensional
sugra action, when compactifying on $S_3 \times S_1$. This,
however, does not generally guarantee consistency of the KK
truncation. Since we get the ansatz of Chamseddine and Sabra as a
limiting case of an ansatz whose consistency was rigorously
proven, we conclude that their ansatz is also consistent.

We begin with $S_4$ written as before as a foliation of $S_3$ with
latitude coordinate $\xi$. Then, we redefine
\be
\xi=a\xi ' \ee and take the limit $a\rightarrow 0^+$ which will
yield an unrestricted $\xi '$, taking values on the real line. 
Then $\tilde\Delta \rightarrow X$, and taking the limit
$a\rightarrow 0$ in (\ref{spl}), we get
\be
ds_{11}^2=X^{1/3} ds_{7}^2 + 2 a^2 g^{-2} X^{10/3} d\xi
`^2+\frac{1}{2} g^{-2}X^{-5/3}\sum_i (\sigma^i -gA_{(1)}^i)^2 
\ee
We also redefine the fields and the coupling constant 
\bea X&=&X'
a^{-2/5}\nonumber\\ g&=&g' a^{2/5}\nonumber\\ A_{(1)}&=&A_{(1)}'
a^{-2/5} 
\eea in order to produce only an overall dependence on
$a$ in $ds_{11}^2$
\be
ds_{11}^2=a^{-2/15}\left(X'^{1/3} ds_7^2+2g'^{-2} X'^{10/3}
{d\xi'}^2+ \frac{1}{2} g'^{-2}X'^{-5/3}\sum_i (\sigma^i
-gA_{(1)}^i)^2\right) \label{scs} 
\ee 
At this moment, noticing that the metric has a Killing vector 
$\frac{\partial}{\partial\xi'}$ we decide to reinterpret $\xi '$ 
as an angular coordinate on $S_1$. With a similar analysis for 
$F_{(4),11}$ given in (\ref{f1}), and with the redefinition
\be
F_{(4),7}=F_{(4),7}' a^{4/5} 
\ee 
we get
\bea
&&F_{(4),11}=a^{-1/5}\left(\frac{g'^{-3}}{\sqrt 2} d\xi
'\wedge\epsilon_{(3)} +\sqrt{2} g'^{-1}X'^4 \star F_{(4),7}'\wedge
d\xi'\right.\nonumber\\
&&-\left.\frac{1}{\sqrt 2} g'^{-2}F_{(2)}'^i \wedge d\xi' \wedge
h^i\right)\nonumber\\&&=a^{-1/5}\frac{g'^{-1}}{\sqrt{2}}d\xi '
\wedge \left(g'^{-2}\epsilon_{(3)}-2X'^4*F_{(4),7}'-g'^{-1}
F_{(2)}'^i\wedge h^i\right)
\label{fcs} 
\eea 
To get rid of the $a$ dependence in the
metric and $F_{(4),11}$ ansatze we exploit the scaling symmetry of
the 11-dimensional sugra equations of motion \cite{cllp,clps} 
\bea
&&ds_{11}^2\rightarrow ds_{11}k^2\\ &&F_{(4),11}\rightarrow
F_{(4),11} k^3 
\eea 
The relation between the scalar field $X'$ and
the dilaton of \cite {cs} is $X'=\exp{4\hat\phi/5}$.
Substituting it in (\ref{scs}, \ref{fcs}) and we are led to
the ansatz of Chamseddine and Sabra
\bea
ds_{11}^2&=&e^{4\hat\phi/15}ds_{7}^2+\frac{1}{4}
e^{-4\hat\phi/3}(\sigma^i-\sqrt{2}A^i_{(1)})^2
+e^{8\hat\phi/3}d\xi^2\nonumber\\
F_{(4),11}&=&d\xi\wedge \left[ dB_{(2)}-(A^i_{(1)}\wedge dA^i_{(1)}
+\frac{1}{3} \epsilon
_{ijk}A^i_{(1)}\wedge A^j_{(1)}\wedge A^k_{(1)})\right.\nonumber\\
&&\left.-\frac{1}{\sqrt{2}}dA^i_{(1)}\wedge \sigma^i +\frac{1}{4}
\epsilon_{ijk}
A^i_{(1)}\wedge\sigma^j\wedge\sigma^k +\frac{1}{12\sqrt{2}}\epsilon_{ijk}
\sigma^i\wedge\sigma^j\wedge\sigma^k
\right]\nonumber\\
\eea
where the two form $B_{(2)}$ is related to the three form $A_{(3),7}'$ through
a duality transformation, and to reach the conventions of Chamseddine and 
Sabra, we set $g'=\sqrt{2}$. 
\\
\item{graviton, two abelian gauge fields $A_{(1)}^i$ and two scalars $X_i$,
i=1,2}\\
This model was discussed in \cite{cdh} and it allows to interpret 2-charge
AdS black holes as the decoupling limit of rotating M2-branes. The ansatz
for a consistent bosonic truncation of $d=11$ sugra to this subset of bosonic
fields reads:
\bea
ds_{11}&=&\tilde\Delta^{1/3}ds_7^2+g^-2\tilde\Delta^{-2/3}\left(X_0^{-1} 
d\mu_0^2+\sum_{i=1}^{2}X_i^{-1}(d\mu_i^2+\mu_i^2(d\phi+gA_{(1)}^i)^2)\right)
\nonumber\\
\hat{\star} F_{(4),11}&=&2g\sum_{i=0}^{2}(X_i^2\mu_i^2-\tilde\Delta X_i) 
\epsilon_{(7)} +g\tilde\Delta X_0\epsilon_{(7)}+\frac{1}{2g}\sum_{i=0}^{2}
X_\i^{-1}\star dX_i\wedge d(\mu_i^2)\nonumber\\
&&+\frac{1}{2g^2}\sum_{i=1}^{2}X_i^{-2} d(\mu_i^2)\wedge (d\phi+gA_{(1)}^i)
\wedge \star F_{(2)}^i
\eea
where $X_0=(X_1 X_2)^{-2}$, $\mu_i, i=0,i,2$, and $\phi_i, i=1,2$ parametrize
the 4-sphere, and $\tilde\Delta=\sum_{i=0}^{2} X_i\mu_i^2$.
The relationship between the Cartesian coordinates $Y^A, A=1,\dots,5$ and
the new variables $\mu_i$ and $\phi_i$ is given by
\bea
Y_5&=&\mu_0\nonumber\\
Y_1&=&\mu_1\sin\phi_1 \;\;\;\;\;\;Y_2=\mu_1\cos\phi_1\nonumber\\
Y_3&=&\mu_2\sin\phi_2\;\;\;\;\;\;Y_4=\mu_2\cos\phi_2
\eea
where the constraint $Y^A Y^A=1$ translates into $\mu_0^2+\mu_1^2+\mu_2^2=1$.
The embedding of the scalar fields in the symmetric matrix $T_{AB}$ of
the maximal $d=7$ gauged sugra is
\be
T_{AB}=diag(X_1,X_1,X_2,X_2,X_0)
\ee
Fianlly, the abelian gauge fields $A_{(1)}^i, i=1,2$ are in the Cartan
subalgebra of $SO(5)_g$:
\bea
\frac{1}{\sqrt2}A_{(1)}^1&=&-\left(B_{(1)}^{12}+B_{(1)}^{34}\right)\nonumber\\
\frac{1}{\sqrt2}A_{(1)}^2&=&B_{(1)}^{12}-B_{(1)}^{34}
\eea
\end{itemize}


\section{The metric ansatz}

We will now analyze the case when we reduce a certain sugra in D
dimensions on a sphere $S_n=SO(n+1)/SO(n)$ and see what we can
learn. In order that the sphere is a background solution for the sugra
of the spontaneous compactification 
type, we need an antisymmetric tensor
$F_{(n)}$, such that this background is of the Freund-Rubin type, i.e.
\be
F_{\mu_1...\mu_n}=\sqrt{\sirc{g}}\epsilon_{\mu_1...\mu_n} 
\ee 
We note that for the compactification of 11d sugra on $AdS_4\times
S_7$, one has a $F_{(4)}$, but we would take the dual of this,
namely an $F_{(7)}$. So, we have to start with an action
\be
S^{(D)}=\int d^Dx \sqrt{g}(R^{(D)} +F_{(n)}^2+...) \label{sD} 
\ee
where ... contain other fields which we put to zero. A first
observation is that this starting point is not the most general.
If one would add a dilaton coupling
\be
S^{(D)}=\int d^Dx \sqrt{g}(R^{(D)} +e^{a(D,n)\phi}F_{(n)}^2+
b(D)(\partial \phi)^2 +...) \ee one would get the most general
starting point for a supergravity action admitting $n-2$ brane
solutions. Then the near-horizon of that p-brane will be an
$AdS_{D-n}\times S_{n}$ background solution of the theory, but the
dilaton would not be constant in that case. We leave the
generalization to this (more interesting) case to future work. But
the action in (\ref{sD}) has been used already in \cite{bbs}, so
it is not without interest to consider it.

We first ask what is the compactification ansatz if in the lower
dimension d=D-n we keep only scalars. We know that the
d-dimensional sugra which is obtained by compactification is a
{\em gauged} sugra, because the sphere $S_n$ has isometry group
$SO(n+1)$, which becomes the gauge group $SO(n+1)_g$ of the
 d dimensional sugra. From the known examples  of gauged 4d sugra, obtained
as 11d sugra on $AdS_4\times S_7$, 7d gauged sugra as 11d sugra on
$AdS_7 \times S_4$, and the less understood case of 5d gauged
sugra as 10d IIB sugra on $AdS_5\times S_5$, we know that the
scalar fields are in a coset G/H, where G is a global isometry of
the ungauged model which is broken down to $SO(n+1)_g$ by the
gauging, and H is a local composite symmetry. The gravitinos of
the gauged sugra are in  a fundamental spinorial representation of
this group H.

Moreover, we know that one way to determine which coset we have is
to count the number of scalar degrees of freedom and the number of
gravitinos and to try to match a coset G/H which does the job. The 
number of scalar degrees of freedom is always the same as for the
torus compactification (going from the torus to the sphere is the
same as going from the ungauged to the gauged model). So the
minimum number of scalar degrees of freedom is given by the
following. The metric will have n(n+1)/2 scalar d.o.f. from the
compact space metric. The antisymmetric tensor will have at least
n scalar d.o.f., coming from the component
$A_{\mu_1...\mu_{n-1}}$, with $\mu_i $ compact space indices. It
could have more, if one of the components with spacetime indices
can be dualized to a scalar. That doesn't happen if
$A_{\alpha_1... \alpha_{n-1}}$ is the d-dimensional dual of a form
$A_{k}$ with $k>0$, i.e. if $d-n-1>0$. This is the case of the
$AdS_7\times S_4$ background, for instance. In such a case, the
coset will be an n(n+1)/2 +n=n(n+3)/2 dimensional manifold. This
is exactly the dimension of $Sl(n+1)/SO(n+1)_c$ $((n+1)^2-1
-n(n+1)/2)$.  Also, we know that the gravitinos multiply Killing
spinors in the linearized KK reduction ansatz (we have something
of the type $\psi_{\alpha I}\eta^I$, as we will see later). That
means that it is safe to assume in that case that the fermions are
in a spinor representation of $SO(n+1)_c$ (because $I$ is a spinor
index for $SO(n+1)_g$).

So, we can say that the scalar coset in the case $d>n+1$ is
determined, namely $Sl(n+1)/SO(n)$. In the other cases, the scalar
manifold will be bigger, but $Sl(n+1)/SO(n)$ will still be a
submanifold, so we will restrict to it in the general case. By a
straightforward extension of the $AdS_7 \times S_4$ case, where
the coset is $Sl(5)/SO(5)$, a coset element will be a matrix
${\Pi_A}^i$, where $A$ is an $SO(n+1)_g$ index and $i$ is an
$SO(n+1)_c$ index. The n(n+3)/2 physical scalars can be described
by the matrix $T^{AB}={\Pi^{-1}_i}^A{\Pi^{-1}_i}^B$, which is
symmetric and has unit determinant. The d-dimensional gauged sugra
will have  gravity+scalar terms:
\be
S^{(d)}=\int d^dx \sqrt{\sirc{g}}(R^{(d)}+P_{\alpha ij}^2 -V(T))
\label{sd} 
\ee 
where $P_{\alpha\,ij}$ is the symmetric part of
$(\Pi^{-1})^A_i{\delta_A}^B \partial_\alpha {\Pi_B}^k
\delta_{kj}$ (where the gauge field was put to zero), and so the
kinetic term in (\ref{sd}) is
$1/8Tr(\partial_{\alpha}T^{-1}\partial^{\alpha}T)$. The scalar
potential is $V(T)=g^2/4(T^2-2T_{AB}^2)$. We will find an ansatz
for the embedding of (\ref{sd}) into (\ref{sD}), and see how to
extend it for the case that
 in (\ref{sd}) we keep also the gauge fields of $SO(n+1)_g$.
We note here that the problem of dimensional reduction to gravity
+ scalars was analyzed also in \cite{cglp}. But the authors of
\cite{cglp} looked only at the cases $AdS_4\times S_7, AdS_7\times
S_4$ and $AdS_5\times S_5$, moreover they restricted themselves to
a diagonal T matrix. That is OK as long as gauge fields don't
couple to scalars, but we want to explore how to generalize to the
case where gauge fields are also nonzero.

Guided by the ansatz in \cite{nvv,nvv2},
 we expect that the metric $g_{\mu\nu}$ on the
compact space has the form
\be
g_{\mu\nu}=\Delta^{\beta}\partial_{\mu}Y^A\partial_{\nu}Y^BT^{-1}_{AB}
\ee where $\Delta $ is $\sqrt{\det g_{\mu\nu}/\det
\sirc{g}_{\mu\nu}}$, with the inverse metric given by
\be
g^{\mu\nu}=\frac{\Delta^{-\beta}}{Y\cdot T\cdot
Y}\partial^{\mu}Y_A
\partial^{\nu}Y_B\left(T^{AB}-\frac{(T\cdot Y)^A(T\cdot Y)^B}{Y\cdot T\cdot Y}
\right) \label{compactul} 
\ee 
But 
\bea &&\det
\tilde{g}_{\mu\nu}\equiv \det T^{-1}_{AB}\partial^{\mu}Y^A
\partial^{\nu}Y^B\equiv \epsilon^{\mu_1...\mu_n}\epsilon^{\mu '_1...\mu '_n}
\tilde{g}_{\mu_1\mu '_1}...\tilde{g}_{\mu_n\mu '_n}\nonumber\\
&&=\epsilon^{\mu_1...\mu_n}\partial_{\mu_1}Y^{A_1}...\partial_{\mu_n}Y^{A_n}
\epsilon^{\nu_1...\nu_n}\partial_{\nu_1}Y^{B_1}...\partial_{\nu_n}Y^{B_n}
T^{-1}_{A_1B_1}...T^{-1}_{A_nB_n}\nonumber\\&& =\sirc{g}
\epsilon^{A_1...A_{n+1}}\epsilon^{B_1...B_{n+1}}Y_{A_{n+1}}Y_{B_{n+1}}
T^{-1}_{A_1B_1}...T^{-1}_{A_nB_n}=\sirc{g}
T^{A_{n+1}B_{n+1}}Y_{A_{n+1}}Y_{B_{n+1}} \nonumber\\
\eea 
where to get from
the second to the third line we have used a simple generalization
of the relations in \cite{nvv2} and the fact that $det T=1$.
 Now we can see that
\be
\Delta^{2-\beta n}=T^{AB}Y_AY_B 
\ee
We take a standard ansatz
for the vielbein, keeping the gauge fields, of the type \bea
&&E_{\alpha}^a=\Delta^{-\frac{1}{d-2}}e_{\alpha}^a(y)\;\;\;,
\;\;\; E_{\alpha}^m=B_{\alpha}^{\mu}E_{\mu}^m\\
&&E_{\mu}^a=0\;\;\;\;\;
,\;\;\;\;E_{\mu}^m=\frac{1}{g}e_{\mu}^m(x,y) 
\eea with $B_{\alpha}^{\mu}=B_{\alpha}^{AB}V^{\mu}_{AB}$, where
$B_{\alpha}^{AB}$ is the $SO(n+1)_g$ gauge field and
$V_{\mu}^{AB}$ is the Killing vector. Then the D-dimensional line
element becomes 
\bea 
ds_D^2&=&\Delta^\alpha
ds_d^2+g^{-2}\Delta^{\beta}T^{-1}_{AB}DY^ADY^B\nonumber\\ &=&
\Delta^{-\frac{2}{d-2}}ds_d^2+g^{-2}\Delta^{\beta}T^{-1}_{AB}DY^ADY^B
\label{ansatzul} 
\eea 
where $DY^A\equiv dY^A+gB^{AB}Y_B$ is a
covariant derivative. We note that the factor
$\Delta^{-\frac{2}{d-2}}$ is necessary to obtain the correct
Einstein action in d dimensions. Next we will determine $\beta$
from the requirement that we recover the correct scalar potential
$V(T)$ after integrating the Einstein action $\int d^DxR^{(D)}$
and the kinetic term of the $n$-form field $\int dx^D F_{(n)}^2$.

First, we compute the contribution coming from the integration of
the Einstein action. In what follows we set (for simplicity reasons) the
Yang-Mills coupling constant to $g=1$. Using the metric ansatz given in
(\ref{ansatzul}) we get 
\bea 
&&\int
\sqrt{g^{(d)}}\sqrt{\sirc{g}^{(n)}}\Delta^{\alpha
d/2+1}d^DxR^{(D)}= \int \sqrt{g^{(d)}}\sqrt{\sirc{g}^{(n)}}
d^Dx\Delta^\alpha \left(R^{(n)}\right.\nonumber\\&&\left.
+\frac{d}{2}\left(\frac{d-3}{2}+\frac{2}
{\alpha}\right)\Delta^{-2\alpha}\partial_{\mu}\Delta^{\alpha}\partial_{\nu}
\Delta^{\alpha}g^{\mu\nu}\right) 
\eea 
where we already used that $\alpha d/2+1=\alpha$. Moreover, if we write
\be
g_{\mu\nu}=\Delta^{\beta}\tilde{g}_{\mu\nu} \label{metricul} 
\ee
we can use the following formula for the relation between the
Ricci scalars of two metrics related by a conformal rescaling as
in (\ref{metricul})
\be
R^{(n)}=\tilde{R}^{(n)}\Delta^{-\beta}-\frac{1}{4}(n-1)
(n-2)g^{\mu\nu}(\partial_{\mu} \ln \Delta^{-\beta})(\partial_{\nu}
ln\Delta^{-\beta}) -(n-1)g^{\mu\nu} D_{\mu}\partial_{\nu}\ln
\Delta^{-\beta} 
\ee 
For our particular metric $\tilde g_{\mu\nu}
=\partial_\mu Y \cdot T \cdot \partial_\nu Y$, the Ricci scalar
has the following expression (independent of the dimensionality of
the compact space): 
\be \tilde{R}^{(n)}=\frac{-2 Y_A Y_B
(T^3)^{AB} +2T Y_AY_B (T^2)^{AB} + Y_AY_BT^{AB} (Tr(T^2)-
T^2)}{(Y_AY_BT^{AB})^2}
\ee

And then we obtain finally 
\bea 
&&\int
\sqrt{g^{(d)}}\sqrt{\sirc{g}^{(n)}}\Delta^{\alpha}d^DxR^{(D)}=
\int \sqrt{g^{(d)}}\sqrt{\sirc{g}^{(n)}}d^Dx
\Delta^{\alpha-\beta}\left\{\tilde{R}^{(n)}-\frac{4}{(2-\beta
n)^2}
\left[\frac{d}{2}\left(\frac{d-3}{2}\right.\right.\right.\nonumber\\
&&\left.\left.\left.+\frac{2}{\alpha}\right)\alpha^2
-\frac{(n-1)(n-2)\beta^2}{4}-\beta(\alpha-1)(n-1)\right]
\left[\frac{(Y\cdot T^2\cdot Y)^2}{(Y\cdot T\cdot
Y)^3} - \frac{Y\cdot T^3 \cdot
Y}{(Y\cdot T\cdot Y)^2}\right]\right\}\nonumber\\{} 
\eea 
and we can already see
that we need to have
\be
\Delta^{\alpha-\beta}=Y\cdot T\cdot Y 
\ee 
in order to obtain terms of
the type $T^2$ in d dimensions. That gives a solution for $\beta$
as
\be
\beta=\frac{2}{n-1}\frac{d-1}{d-2} \ee And now we can write the
full line element as
\be
ds_D^2=(Y\cdot T\cdot Y)^{-\frac{2}{(d-2)(2-\beta n)}}ds_7^2 +(Y\cdot
T\cdot Y) ^{\frac{\beta}{2-\beta n}}T^{-1}_{AB}DY^ADY^B
\label{wow} \ee with $\beta$ found before. We can compare with the
ansatz given in \cite{cglp} for the case of no gauge fields and a
diagonal matrix $T_{AB}=X_A\delta_{AB}$, valid for the
$AdS_4\times S_7, AdS_7\times S_4 $ and $AdS_5\times
S_5$compactifications, namely
\be
ds_D^2=(X_AY_A^2)^{\frac{2}{d-1}}ds_d^2+(X_AY_A^2)^{-\frac{d-3}{d-1}}
(X_A^{-1}dY_A^2) \ee By comparing the two results, we see that we
have the same result if $d(n-3)=3n-5$, an equation which has as
the ONLY solutions ($d=4, n=7$), ($d=7,n=4$) and ($d=n=5$). So we have
the curious fact that the formula derived in \cite{cglp}, although
is written in a general form (with a general $d$), is only valid in
the cases for which it was derived, not for general $d$. For general
$d$ and $n$, one should take the appropriate truncation of our formula
(\ref{wow}).

Finally, we will complete the calculation of the scalar potential.
We can compute the integrals on the sphere using the formulas 
\bea
\int\frac{Y^AY^B}{Y\cdot T\cdot Y}&=&\frac{1}{n+1}T^{-1}_{AB}
\\
\int\frac{Y^AY^BY^CY^D}{(Y\cdot T\cdot
Y)^2}&=&\frac{1}{(n+1)(n+3)} (T^{-1}_{AB}T^{-1}_{CD}\nonumber\\&&
+T^{-1}_{AC}T^{-1}_{BD}+T^{-1}_{AD}T^{-1}_{BC}),\;\;{\rm etc.} 
\eea
From that we derive, for instance
\be
\int\left[\frac{Y\cdot T^3\cdot Y}{Y\cdot T\cdot
Y}-\left(\frac{Y\cdot T^2 \cdot Y}{Y\cdot T \cdot
Y}\right)^2\right]=\frac{1}{n+3}\left(T_{AB}^2
-\frac{T^2}{n+1}\right) 
\ee  
In conclusion, the Einstein term in the D-dimensional action gives the 
following contribution to d-dimensional the scalar potential
\be
(n-1)\int \sqrt{g^{(d)}} d^dx\left[ \frac{d-1}
{d+n-2}\frac{1}{n+3}(T_{AB}^2-\frac{T^2}{n+1}) +\frac{1}{n+1}
(T_{AB}^2-T^2)\right]
\label{eterm}
\ee

We will now turn to the ansatz for the antisymmetric tensor field.
We will again generalize the result in \cite{nvv2}, where the
ansatz for the 3-index tensor in the presence of only scalar
fields is
\be
A_{(3)}=-\frac{1}{6\sqrt{2}}\epsilon_{A_1...A_5}dY^{A_1}\wedge dY^{A_2}
\wedge dY^{A_3}
Y^{A_4}\left(\frac{T\cdot Y}{Y\cdot T\cdot Y}\right)^{A_5} 
\ee 
We write, similarly, an ansatz for the $n-1$ form 
\bea
A_{(n-1)}&=&-a\epsilon_{A_1...A_{n+1}}dY^{A_1}\wedge...\wedge
dY^{A_{n-1}} Y^{A_n}\left(\frac{T\cdot Y}{Y\cdot T\cdot
Y}\right)^{A_{n+1}}+\nonumber\\
&&-\frac{1}{n}\frac{\sirc{D}_{\mu_n}}{\sirc{\Box}}(\epsilon_{\mu_1...\mu_n}
\sqrt{\sirc{g}}) 
\eea 
where we kept only one factor of $T\cdot
Y/(Y\cdot T\cdot Y)$ because we want to have an d-dimensional
action involving only $T^2$ terms (and no $TrT^{-1}$ terms, for
example), completely analog to the case considered in \cite{nvv2}.
Here $a$ is a constant to be fixed. However, if we look then at the
terms with only scalars and no derivative in $F_{(n)}$, we have
\bea 
&&F_{(n)}|_{no\;\partial
T}=(\sqrt{\sirc{g}}\epsilon_{\mu_1...
\mu_n}dx^{\mu_1}...dx^{\mu_n})\left(1\right.\nonumber\\&&\left.+\frac{a}{n}
\left[ \frac{T}{Y\cdot T\cdot Y}-(n+1) -2\left(\frac{Y\cdot
T^2\cdot Y}{(Y\cdot T\cdot Y)^2}-1\right)\right]\right) 
\eea 
and then the term in the action becomes
\be
F_{(n)}^2\Delta^{\alpha}|_{no\;\partial T}=\Delta^{\alpha -2}n!\left[
1-\frac{a(n-1)}{n} +\frac{a}{n}\left(\frac{T}{Y\cdot T\cdot
Y}-2\frac{Y\cdot T^2\cdot Y}{(Y\cdot T\cdot Y)^2}\right)\right]^2
\label{bracket} 
\ee 
But in order to get only $T^2$ terms in the
action, we need that the the power of $\Delta$ is $2(2-\beta n)$,
so that we obtain a factor $(Y\cdot T \cdot Y)^2$ multiplying the
bracket in (\ref{bracket}). That gives again the equation
$d(n-3)=3n-5$, which is solved only by ($d=4,n=7$), ($d=7,n=4$) and
($d=n=5$).

But we can easily see how we can modify the ansatz. Just multiply
$F_{(n) \mu_1...\mu_n}$ by $\Delta^{3-\beta n-\frac{\alpha}{2}}$.
Then $A_{(n-1)}$ will be 
\bea 
A_{(n-1)}&=&-\Delta^{3-\beta
n-\frac{\alpha}{2}}
a\epsilon_{A_1...A_{n+1}}dY^{A_1}\wedge...\wedge dY^{A_{n-1}}
Y^{A_n}\left(\frac{T\cdot Y}{Y\cdot T\cdot
Y}\right)^{A_{n+1}}+\nonumber\\
&&-\frac{1}{n}\frac{\sirc{D}_{\mu_n}}{\sirc{\Box}}(\epsilon_{\mu_1...\mu_n}
\sqrt{\sirc{g}}\Delta^{3-\beta n-\frac{\alpha}{2}}) 
\eea 
Notice that we could have put $\Delta^{2-\beta n-\frac{\alpha}{2}}$
outside $\frac{\sirc{D}_{\mu_n}}{\sirc{\Box}}$, but that would
generate an extra term in the scalar potential. Now the term in the action
becomes
\bea
F_{(n)}^2\Delta^{\alpha}|_{no\;\partial T}&=&n!\int (Y\cdot T\cdot Y)\left[
1-\frac{a(n-1)}{n} +\frac{a}{n}\left(\frac{T}{Y\cdot T\cdot
Y}-2\frac{Y\cdot T^2\cdot Y}{(Y\cdot T\cdot Y)^2}\right)\right.\nonumber\\&&
\left.+2a\left(1-\frac{\alpha -2}{2(2-\beta n)}\right)\left(1-\frac{Y\cdot T^2
\cdot Y}{(Y\cdot T\cdot Y)^2}\right)\right]^2
\label{fterm}
\eea 
and by integration we get a combination of $T_{AB}^2$ and $T^2$ terms, with a 
free parameter (a). Then a is fixed by requiring that in the sum of 
(\ref{fterm}) and (\ref{eterm}) the relative coefficient of $T_{AB}^2$ 
and $T^2$ is $-2$ (the overall coeffiecient depends on the coupling constant
g). It is not clear how one gets the correct kinetic terms fot T, since 
$\partial_{\alpha}A_{(n-1)\beta\gamma\delta} $ will contain 
$\frac{\sirc{D}_{\mu_n}}{\sirc{\Box}}$ terms. When integrated on the sphere,
they should give contributions to the $P_{\alpha ij}^2$ kinetic terms too.

In the remainder of this section we will look at applications of
our metric ansatz (\ref{wow}). Let us see that it can be applied
for the $AdS_7\times S_4$ case. The metric deduced by de Wit and
Nicolai \cite{dwnw} is 
\bea 
\Delta^{-1}g^{\mu\nu}&=&\frac{1}{8}
V^{\mu IJ}V^{\nu KL}{w_{i'j'}}^{IJ} {w^{i'j'}}_{KL} \nonumber\\
{w_{ij}}^{IJ}&=&={u_{ij}}^{IJ}+v_{ijIJ} \;\; ;\;\; {w^{ij}}_{KL}=
{u^{ij}}_{KL}+v^{ijKL}\label{dewitul}
\eea
where I,J,... are spinorial $SO(8)_g$
 indices, and i',j',... are  $SU(8)_c$ indices, both [i'j'] and [IJ] are
antisymmetrized, and  $u$ and $v$ together form a representation of
$E_7$. Then one can define 
\bea 
V^{\mu AB}&=&(\Gamma^{AB})^{IJ}V^{\mu}_{IJ}\\
{w_{ij}}^{AB}&=&\frac{1}{64}(\Gamma^{AB})_{IJ}{w_{ij}}^{IJ} 
\eea 
where A,B,... are vector $SO(8)_g$ indices, and
$V_{\mu}^{AB}=Y^A\partial_{\mu} Y^B$ and we used that $(\Gamma^{AB})_{IJ}
(\Gamma_{AB})^{KL}=16\delta_{IJ}^{KL}$. 
Then if one restricts the
scalar coset from $E_7/SU(8)_c$ to $Sl(8)/SO(8)_c$, with $SO(8)_g$
embedded in $Sl(8)$, the matrix $w$ will reduce to
\be
{w_{ij}}^{AB}={\Pi^{-1}_{[i}}^A{\Pi^{-1}_{j]}}^B \ee and
\be
{w_{i'j'}}^{AB} {w^{i'j'}}_{CD}=T^{[A}_{[C}T^{B]}_{D]} 
\ee 
and then we recover our general formula (\ref{compactul}).

As a small application, we note that now we can write full line
element as
\be
ds_{11}^2=\Delta^{-1}ds_7^2+\Delta^{-1}M_{AB}DY^ADY^B \label{linel}
\ee where $M_{AB}$ is the 'inverse' of $\tilde{M}^{AB}=
{w_{i'j'}}^{AC} {w^{i'j'}}_{BD}Y_CY^D$, i.e.
\be
\tilde{M}^{AC}M_{BC}=\delta^A_B+(Y_B {\rm terms}) \label{inverse}
\ee 
Note that $\tilde{M}$ has no inverse in the usual sense,
because it has an eigenvector with zero eigenvalue, namely $Y_A$,
but we only need (\ref{inverse}) to get (\ref{linel}).

A new application for our method is an  ansatz for the metric in
the $AdS_5\times S_5$ case. We guess it by the inverse procedure
to the one described for the $AdS_4\times S_7$ case. Namely, we
replace ${\Pi^{-1}_{[i}}^A{\Pi^{-1}_{j]}}^B$ by the projection of
the  $E_6/USp(8)_c$ scalar coset vielbein onto the {\bf 15}
representation of $SO(6)_g$,
\be
{\Pi^{-1}_{[i}}^A{\Pi^{-1}_{j]}}^B\rightarrow {(\Pi^{-1}({\bf
15}))_{ab}}^{AB}\equiv (\Gamma^{AB})_{\alpha\beta} {P({\bf
15})^{\alpha\beta}}_{\gamma\delta}{(\Pi^{-1})_{ab}}^{\gamma\delta}
\ee 
In this way we arrive at the ansatz
\be
\Delta^{-2/3}g^{\mu\nu}=V^{\mu}_{AB}V^{\nu}_{CD} {(\Pi^{-1}({\bf
15}))_{ab}}^{AB}{(\Pi^{-1}({\bf 15}))_{ab}}^{CD} \label{5dmetric}
\ee 
as a natural extension of our particular ansatz
\be
\Delta^{-2/3}g^{\mu\nu}=V^{\mu}_{AB}V^{\nu}_{CD}T_{AC}T_{BD} \ee
We will give another argument for (\ref{5dmetric}) in the
following section, by first computing the vielbein.

We note that we can again put the 10d line element in a form
similar to (\ref{linel}), namely:
\be
ds_{10}^2=\Delta^{-\frac{2}{d-2}}ds_5^2+\Delta^{-2/3}N_{AB}DY^ADY^B
\ee where $N_{AB}$ is the 'inverse' of $\tilde{N}_{AB}=
{(\Pi^{-1}({\bf 15}))_{ab}}^{AC}Y_C{(\Pi^{-1}({\bf
15}))_{ab}}^{BD}Y_D$, i.e.
\be
\tilde{N}^{AC}N_{BC}=\delta^A_B+(Y_B {\rm terms}) \ee


\section{The fermion fields and vielbein ansatz}

In this section we will try to give a general discussion of the
fermionic fields and the vielbein in the case of compactifications
on spheres.

For the vielbein, we already gave a general ansatz, namely \bea
&&E_{\alpha}^a=\Delta^{-\frac{1}{d-2}}e_{\alpha}^a\;\;\;, \;\;\;
E_{\alpha}^m=B_{\alpha}^ {\mu}E_{\mu}^m\\ &&E_{\mu}^a=0\;\;\;\;\;
,\;\;\;\;E_{\mu}^m \eea with
$B_{\alpha}^{\mu}=B_{\alpha}^{AB}V^{\mu}_{AB}$, the only thing
that remains to give is an ansatz for the compact space vielbein,
$E_{\mu}^m$. For that, we need first an ansatz for the fermions.

In a supergravity theory, we always have a gravitino field. By
dimensional reduction, we obtain a set of gravitinos in d
dimensions, transforming in the fundamental representation under
the composite symmetry of the gauged sugra. For example, in the
$AdS_7\times S_4$ case, we have gravitinos in the {\bf 4} of
$SO(5)_c=USp(4)_c$. For the $AdS_4\times S_7$ case we have
gravitinos in the {\bf 8} of $SU(8)_c$, and
 in the $AdS_5\times S_5$ case, the gravitinos are in the {\bf 8} of
$USp(8)_c$. Throughout most of the discussion we will restrict
ourselves to a $SO(n+1)_c$ composite symmetry group ($SO(5)_c,
SO(8)_c\in SU(8)_c, SO(6)_c\in USp(8)_c$).

However, the usual (linearized) ansatz for the gravitinos involves
a Killing spinor ($\Psi_{\alpha}(y,x)=\psi_{\alpha
I}(y)\eta^I(x)$), which has an $SO(n+1)_g$ spinor index of the
gauge field. That tells us that there should exits a
 matrix ${U^{I'}}_I$, with $I'$ a composite group index, interpolating
between the gravitinos and the Killing spinor.

The result we obtained in 7d in \cite{nvv,nvv2} is \bea
\Psi_a&=&\Delta^{1/10}(\gamma_5)^{1/2}\psi_a
-\frac{1}{5}\tau_a\gamma_5 \gamma^m
\Delta^{1/10}(\gamma_5)^{-1/2}\psi_m\label{useful}\\
\Psi_m&=&\Delta^{1/10}(\gamma_5)^{-1/2}\psi_m\label{useful2}\\
\varepsilon&=&\Delta^{-1/10}(\gamma_5)^{1/2}\epsilon,
\bar{\varepsilon}=\Delta^{-1/10}\bar{\epsilon}(\gamma_5)^{1/2}
\nonumber\\ \psi_{\alpha}(y,x)&=& \psi_{\alpha
I'}(y)U^{I'}\;_I(y,x)\eta^I(x) \label{psia}\;\; where\;\;
\psi_a=e_a^{\alpha}(y)\psi_{\alpha}(y,x)\\ \psi_m(y,x)&=&
\lambda_{J'K'L'}(y)U^{J'}\;_J(y,x)U^{K'}\;_K(y,x)U^{L'}\;_L(y,x)
\eta_m^{JKL}(x) \label{psim}
\\ \epsilon (y,x)&=&\epsilon_{I'}(y)U^{I'}\;_I\eta^I(x)\label{epsi}
\eea We see that we needed a certain rotation to diagonalize the
7dimensional gravitino and spin 1/2 kinetic operators. In general,
we expect that the rotation is more complicated, such that both
$\Psi_a $ and $\Psi_m$ depend on both gravitino and spin 1/2
fields in d dimensions. Moreover, in general, the D dimensional
sugra itself has spin 1/2 fields, so the rotation involved is
probably more general than the one considered here. We will not
have more to say about that, we will just note that the advertised
matrix ${U^{I'}}_I$ appears in the ansatz for all the fermions.

One possible interpretation is that ${U^{I'}}_I$ gives a (field-
and spherical harmonic-dependent) composite rotation. Indeed,  U
is a matrix in the coset $H_c/SO(n+1)_g$ (for instance because it
has one index transforming under $H_c$ and one transforming under
$SO(n+1)_g$), so it can be interpreted as a field dependent $H_c$
rotation. We note that if we break $H_c$ to $SO(n+1)_c$ (we will
do it for most of this section), then U becomes simply an
$SO(n+1)$ matrix in the spinor representation. Since the vielbein
couples to the fermions, we should expect that $E_{\mu}^m$ depends
on U also, and we shall find exactly that. We note here that restricting 
U to be and $SO(n+1)$ matrix in the spinor representation means also that the 
fermions are restricted to lie in a spinor representation of $SO(n+1)$. For 
a concrete example, for the $AdS_4\times S_7$ case that would mean that 
the fermions, which are in a representation of SU(8) are restricted to 
transform only under an SO(8) representation.

We can however still say the following about the fermions. We take
$\Gamma_a= \tau_a\times \gamma$, as the dimensional reduction of
the gamma matrices from D to d dimensions ($\Gamma_M $ to
$\tau_a$), where $\gamma$ is one for odd-dimensional spheres and
$\gamma_{2n+1}$ for even-dimensional spheres. If both d and n are odd, the
gamma matrix split involves an extra $\sigma$ matrix, e.g. in $AdS_5\times
S_5$ we have $\Gamma_a=\tau_a\times \id\times \sigma_1$ (and $\Gamma_m
=\id\times \gamma_m\times (-\sigma _2)$). We will examine the $AdS_5\times
S_5$ case separately.
From the requirement
that the D dimensional gravitino action reduces to the d
dimensional one, more precisely that
\be
\Delta^{-\frac{2}{d-2}}\Psi_a\Gamma^{ab\alpha}\partial_{\alpha}\Psi_b
=\psi_{aI'}\tau^{ab\alpha}\partial_{\alpha}\psi_b^{I'}+... \ee (on
the left-hand-side indices are flattened with the D-dimensional
vielbein and on the right-hand-side with the d-dimensional one),
we get the ansatz
\be
\Psi_a(y,x)=\Delta^{\frac{1}{2(d-2)}}(\gamma)^p\psi_{a
I'}{U^{I'}}_I(y,x) \eta^I(x)+{\rm spin\;\;1/2\;\;terms} \ee where
again $\Psi_a$ is flattened with the D dimensional vielbein and
$\psi_a$ with the d-dimensional one. Moreover, we can derive the
ansatz for the susy parameter,
\be
\varepsilon(y,x)=\Delta^{-\frac{1}{2(d-2)}}(\gamma)^p\epsilon_{I'}{U^{I'}}_I
\eta^I \ee Using this fermionic ans\"atze, we will derive the
vielbein ansatz and then the relation that the matrix U has to
satisfy.

The vielbein ansatz is easily derived from the susy law of the
gauge fields. We have in general a susy law of the type
\be
\delta B_{\alpha}^{AB}={(\Pi^{-1})_{I'J'}}^{AB}(\bar{\epsilon}
^{I'}\psi_{\alpha} ^{J'}){\rm + spin \;1/2\;\; terms} \ee That is
so because a boson transforms into $\epsilon$ times a fermion, and
by matching indices, it is easy to see that we need a bosonic
matrix in a coset involving the gauge group (via the A,B indices)
and the composite group (via the $I',J'$ indices). The only one
available is a matrix ${(\Pi^{-1})_{I'J'}}^{AB}$ which is either
the scalar coset vielbein, or some product of such vielbeins. We
will treat separately the cases of $AdS_4\times S_7, AdS_7\times
S_4$ and $AdS_5\times S_5$, but in the case that we restrict the
scalar manifold to the
 coset $Sl(n+1)/SO(n+1)$ (in particular, for
$AdS_7\times S_4$ there is no restriction made), we know that the
susy law takes the form
\be
\delta
B_{\alpha}^{AB}=c{\Pi^{-1}_i}^A{\Pi^{-1}_j}^B(\gamma^{ij})_{I'J'}
(\bar{\epsilon}^{I'}\psi_{\alpha}^{J'}){\rm + spin \;1/2\;\;
terms} \ee where c is a normalization constant. On the other hand,
we can obtain the variation of $B_{\alpha}^{\mu}$ starting from D
dimensions, in a manner entirely analog to the one used in
\cite{nvv2}. Namely, by assuming only that in D dimensions we have
the following susy law for the vielbein
\be
\delta
E^M_{\Lambda}=\frac{1}{2}\bar{\varepsilon}\Gamma^M\Psi_{\Lambda}
\ee and using the fact that in the gauge $E_{\mu}^a=0$ we have a
compensating off-diagonal Lorentz rotation
${\Omega_m}^a=-{\Omega^a}_m=-\frac{1}{2}
\bar{\epsilon}\Gamma^a\Psi_m$, we get the variation of
$B_{\alpha}^{\mu}$, \bea \delta
B_{\alpha}^{\mu}&=&\delta_{susy}(E_{\alpha}^mE_m^{\mu})
+E_{\alpha}^m(E_a^{\mu}{\Omega^a}_m)+(E_{\alpha}^a{\Omega_a}^mE^{\mu}_m
\nonumber\\ &=&\frac{1}{2}E_{\mu}^mE_{\alpha}^a\bar{\varepsilon}
[\Gamma_a\Psi^m+\Gamma^m\Psi_a] 
\eea 
Now we can substitute the
gravitino ansatz and (in the hypothesis that there is no
d-dimensional gravitino contribution in $\Psi_m$), we get
\be
\frac{s}{2}\Delta^{-\frac{1}{d-2}}E^{\mu}_m[\epsilon\gamma^m\psi_{\alpha}
+{\rm spin\;\;1/2\;\;terms}] 
\ee 
where s is a phase coming from
the commutation relation of $(\gamma)^p$ with $\gamma^m$. By
comparing with the d-dimensional result, we get
\be
\Delta^{-\frac{1}{d-2}}E^{\mu}_mV^{m,IJ}{U^{I'}}_I{U^{J'}}_J=\frac{2c}{s}
{(\Pi^{-1})_i}^A{(\Pi^{-1})_j}^BV^{\mu}_{AB}(\gamma^{ij})^{I'J'}\label{viel}
\ee 
and therefore we obtain an ansatz for the vielbein,
\be
\Delta^{-\frac{1}{d-2}}E^{\mu}_m=\frac{2ce}{s}
{(\Pi^{-1})_i}^A{(\Pi^{-1})_j}^BV^{\mu}_{AB}Tr(\gamma^{ij}U V_n
U^{-1}) \label{viel2} 
\ee 
Here e is defined by
$(UV^mU^T)^{IJ}(UV_nU^T)_{IJ}=e\delta_n^m$. Squaring the result
for the vielbein we get back the metric in (\ref{compactul}), 
provided $4c^2=s^2e$.

We also get a self-consistency condition on the matrix U, by
plugging (\ref{viel2}) back into (\ref{viel}),
\be
e{(\Pi^{-1})_i}^A{(\Pi^{-1})_j}^BV^{\mu}_{AB}Tr(\gamma^{ij}U V_m
U^{-1}) V^{m,IJ}{U^{I'}}_I{U^{J'}}_J=
{(\Pi^{-1})_i}^A{(\Pi^{-1})_j}^BV^{\mu}_{AB}(\gamma^{ij})^{I'J'}
\label{consis} 
\ee 
This equation is the most general condition on
the matrix U. In \cite{nvv2}, we found the equation (in the $AdS_7\times 
S_4$ case)
\be
UY\llap/=v\llap/U,\;\;v_i=\Delta^{-3/5}
{(\Pi^{-1})_i}^AY_A\label{uyvu} 
\ee 
and showed that it implies
(\ref{consis}). But the reverse is also true: (\ref{consis})
implies (\ref{uyvu}). If one multiplies (\ref{consis}) by
$(\gamma_{kl})_{I'J}$, one gets on the left-hand side:
\be
Tr(UY\llap/U^{-1}\gamma_{kl}UY\llap/
U^{-1}\gamma^i\gamma^j){(\Pi^{-1})_i}^A
Y_A{(\Pi^{-1})_j}^BC_{\mu}^B 
\ee 
But $UY\llap/ U^{-1}$ will be
equal to $v\llap/$, where v is a unit vector, just because U is a
SO(n+1) matrix in the spinor representation. To determine v, we
impose that we recover the right-hand side and obtain
\be
v_i=\Delta^{-3/5} {(\Pi^{-1})_i}^AY_A 
\ee 
as desired. So we can say
that (\ref{consis}) is the most general equation on U, but it can
be brought to the nicer form (\ref{uyvu}).

Although we look here at the case $AdS_7\times S_4$, this is valid for all
(d,n), the only thing we used here was that $Tr M\gamma_A Tr N\gamma_A
\propto Tr MN$ (\ref{compln}), under the assumption $Tr M=Tr N=0$, and that 
$UY\llap/ U^{-1}=v\llap/ $ for some $v$. Then we get in general
\be
v_i=\Delta^{\beta n/2-1}{(\Pi^{-1})_i}^AY_A
\ee

We will now finally turn to the specific cases of the $AdS_4\times
S_7$ and the $AdS_5\times S_5$ compactifications.

For the $AdS_4\times S_7$ compactification, the full vielbein
ansatz (with no restriction to the Sl(8)/SO(8) scalar coset) was
implicitly derived in \cite{dwnw}, by the equation
\be
\frac{\sqrt{2}}{8}i\Delta^{-1/2}E^{\mu}_m(U V^m U^T)_{ij}
={w_{ij}}^{IJ}V^{\mu}_{IJ} 
\label{viel4}
\ee 
and we can derive the vielbein and
the equation for U from it. The vielbein is
\be
\frac{\sqrt{2}}{8}i\Delta^{-1/2}E^{\mu}_m
=-\frac{1}{16}{w_{ij}}^{IJ}V^{\mu}_{IJ}(U V^m U^T)_{ij} 
\ee 
(we used here that $(UV^m U^{-1})_{ij}(UV^n U^{-1})^{ij}=-16 \delta_m^n$)
and the equation for U (the selfconsistency equation) is
\be
-\frac{1}{16}{w_{kl}}^{IJ}V^{\mu}_{IJ}(U V^m U^T)_{kl}(U V^m U^T)_{ij}
={w_{ij}}^{IJ}V^{\mu}_{IJ} 
\ee 
We note that now we can't find a
nicer form for the U equation, because the vielbein $w$ doesn't
factorize into two, as it does if we restrict to the $Sl(8)/SO(8)$
coset. We also note that from the vielbein (\ref{viel4}) one can rederive the 
metric (\ref{dewitul}).

For the $AdS_5\times S_5$ case, the ansatz for the vielbein
 can be obtained easily via the
general procedure outlined before. From the 10d transformation
law, we get
\be
\delta B_{\alpha}^{\mu}
\frac{1}{2}\Delta^{-1/3}E^{\mu}_m[\varepsilon
\gamma^m\psi_{\alpha}+ {\rm spin\;\;1/2\;\;terms}] \ee whereas the
5d susy law is
\be
\delta B_{\alpha}^{AB}=2i {(\Pi^{-1}({\bf
15}))_{ab}}^{AB}(\bar{\epsilon}^a \psi_{\mu}^b+{\rm spin\;\;1/2
\;\;terms}) 
\ee 
implying an equation for the vielbein,
\be
\Delta^{-1/3}E^{\mu}_m V^{m, IJ}{U^a}_I{U^b}_J =4i{(\Pi^{-1}({\bf
15}))_{ab}}^{AB}V^{\mu}_{AB} \label{ads5v}
\ee 
Here ${U^a}_I$ is a matrix in the
coset $USp(8)/SO(6)_g$. Using $(UV^m U^{-1})_{ab}(UV_n U^{-1})^{ab}
=-4\delta_m^n$ we can square (\ref{ads5v}) to get back (\ref{5dmetric}), as
promised. The vielbein then is
\be
\Delta^{-1/3}E^{\mu}_m=-i{(\Pi^{-1}({\bf
15}))_{ab}}^{AB}V^{\mu}_{AB} V^{m, IJ}{U^a}_I{U^b}_J 
\ee 
and the selfconsistency equation (the equation for the matrix U) is
\be
-\frac{1}{4}{(\Pi^{-1}({\bf 15}))_{ab}}^{AB}V^{\mu}_{AB} V^{m,
IJ}{U^a}_I{U^b}_J V^{m, IJ}{U^a}_I{U^b}_J ={(\Pi^{-1}({\bf
15}))_{ab}}^{AB}V^{\mu}_{AB}  \label{5dueq} 
\ee
A comment is in order for the fermionic ansatz for the $AdS_5\times S_5$
case: in 10d we already have two gravitinos, which one can take to 
form a complex gravitino. Then the KK reduction will have the same form 
as the general one, but now the Killing spinors are in the spinorial 
representation of $SO(6)_g$, or the fundamental representation of $SU(4)$, 
which is a 4d complex representation. The 5d gravitinos are in the fundamental
of $USp(8)_c$, which is a 8d real representation, so one has a choice of 
writing the U matrix as an 8 by 8 real matrix, or a 4 by 4 complex one.


\section{Conclusions and discussions}

We have shown that the ansatz in \cite{nvv2} can be used for a variety of 
purposes. By further truncating the number of fields, we were able to 
reproduce the ans\"atze of \cite{lp,cglp,cdh,cs,cllp}. That is to be expected
since the KK truncation of 11d sugra on $AdS_7\times S_4$ in \cite{nvv2}
is a {\em consistent} one (i.e. satisfies the higher dimensional 
equations of motion and susy laws), as are the truncations found in 
 \cite{lp,cglp,cdh,cs,cllp}. As a byproduct of our analysis, we find a 
lagrangean formulation of the ${\cal N}=2$ model, which was missing 
in the literature.
This ${\cal N}=2$ model was described until now only through 
equations of motion,
supplemented with the self-duality constraint. The lagrangean which
describes the bosonic sector of the N=2 model reads:
\bea
\sqrt{g_7^{-1}}L_{7d, {\cal N}=2}&=&R+g^2(2X^{-3}+2X^2-X^{-8})+5\partial_
{\alpha}X^{-1}\partial_{\alpha}X+\frac{g^2}{2} X^{-4}S_{\alpha\beta\gamma}
S^{\alpha\beta\gamma}\nonumber\\
&&-\frac{1}{4}X^{-2}F_{\alpha\beta}^i F^{\alpha\beta \;i}
+\epsilon^{\alpha\beta\gamma\delta\epsilon\eta\zeta} \sqrt{g_7^{-1}}
\left(\frac{g}{24\sqrt 2}S_{\alpha\beta\gamma}F_{\delta\epsilon\eta\zeta}
\right.\nonumber\\&&+\left.
(-\frac{1}{48\sqrt 2 g}\omega_{\alpha\beta\gamma} +\frac{i}{8\sqrt{3}}
S_{\alpha\beta\gamma})F_{\delta\epsilon}^i F_{\eta\zeta}^i\right)	
\eea
where the three form $A_{(3)}$ used so far in the ${\cal N}=2$ model is 
a linear combination of the three form $S_{(3)}$ of the maximal gauged 
sugra model and the Chern-Simons three form $\omega_{(3)}$ (\ref{comb}). 

Then we have found that if we make the KK reduction on a sphere $S_n$ 
of a theory involving
gravity and an antisymmetric tensor $F_{(n)}$ to a theory involving
gravity and  scalars, the scalar coset has a submanifold $Sl(n+1)/SO(n+1)_c$,
and the ansatz for the line element is
\be
ds_D^2=(Y\cdot T\cdot Y)^{-\frac{2}{(d-2)(2-\beta n)}}ds_7^2 +(Y\cdot T\cdot Y)
^{\frac{\beta}{2-\beta n}}T^{-1}_{AB}DY^ADY^B
\ee
where $\beta=\frac{2}{n-1}\frac{d-1}{d-2}$ and $DY^A=dY^A+gB^{AB}Y_B$ is a 
covariant derivative. For the antisymmetric tensor, 
the ansatz for the dependence on the coset scalars is
\bea
A_{(n-1)}&=&-\Delta^{3-\beta n-\frac{\alpha}{2}}
a\epsilon_{A_1...A_{n+1}}dY^{A_1}\wedge...\wedge dY^{A_{n-1}}
Y^{A_n}\left(\frac{T\cdot Y}{Y\cdot T\cdot Y}\right)^{A_{n+1}}+\nonumber\\
&&-\frac{1}{n}\frac{\sirc{D}_{\mu_n}}{\sirc{\Box}}(\epsilon_{\mu_1...\mu_n}
\sqrt{\sirc{g}}\Delta^{3-\beta n-\frac{\alpha}{2}})\\
F_{(n)}|_{no\;\partial T}&=&\Delta^{3-\beta n-\frac{\alpha}{2}}
(\sqrt{\sirc{g}}\epsilon_{\mu_1...
\mu_n}dx^{\mu_1}...dx^{\mu_n})\left(1+\right.\nonumber\\&&\left.+\frac{a}{n}
\left[ \frac{T}{Y\cdot T\cdot Y}-(n+1)
-2\left(\frac{Y\cdot T^2\cdot Y}{(Y\cdot T\cdot Y)^2}-1\right)\right]\right)
\nonumber\\&&
+2a\left(1-\frac{\alpha -2}{2(2-\beta n)}\right)\left(1-\frac{Y\cdot T^2
\cdot Y}{(Y\cdot T\cdot Y)^2}\right)
\eea
and we see that by making derivatives covariant we can couple to gauge fields
as in the metric. But there are  other terms involving the Yang-Mills field
strength  in the 
ansatz for the antisymmetric tensor, as we can easily see in the $AdS_7
\times S_4$ example. By imposing Bianchi identity on $F_{(n)}$ we can
presumably generate these terms. However, it is possible to miss a 
separately closed form, gauge-invariant, and which vanishes when the scalar 
fields are set to zero. So, the ansatz for the antisymmetric tensor
field strength generated by this method should be checked in the action
too, to verify its completeness.  

Moreover, we note that the $\partial T$ terms in $F_{(n)}$ contain 
$1/\Box_x$ terms, and so we were not able to reproduce the scalar field 
kinetic terms in the d dimensional action,
except in the cases of $AdS_4\times S_7$, $AdS_7\times S_4$ and 
$AdS_5\times S_5$, when these $1/\Box_x$ terms disappear already in $F_{(n)}$. 

In the $AdS_5\times S_5$ case, we got a line element,
\be
ds_{10}^2=\Delta^{-2/3}(ds_5^2+N_{AB}DY^ADY^B)
\label{me}
\ee
where $N_{AB}$ is the 'inverse' of $\tilde{N}_{AB}=
{(\Pi^{-1}({\bf 15}))_{ab}}^{AC}Y_C{(\Pi^{-1}({\bf 15}))_{ab}}^{BD}Y_D$, i.e.
\be
\tilde{N}^{AC}N_{BC}=\delta^A_D+(Y_D {\rm terms})
\ee
One can now use this to lift any solution of gauged sugra to 11 dimensions, 
at least for the metric.

We were able to obtain the vielbein too, by first fixing a small part of the 
fermion ansatz, namely the relation between the D dimensional and the d 
dimensional gravitinos. We found that also in the general case we need a 
matrix ${U^{I'}}_I$ interpolating between the composite symmetry group indices 
of the gravitino and the gauge group spinor indices of the Killing spinors. 
This matrix apppears in the ansatz for the vielbein, which for the 
case of the $Sl(n+1)/SO(n+1)$ coset reads
\be
\Delta^{-\frac{1}{d-2}}E^{\mu}_m=\frac{2ce}{s}
{(\Pi^{-1})_i}^A{(\Pi^{-1})_j}^BV^{\mu}_{AB}Tr(\gamma^{ij}U V_n U^{-1})
\ee
In this case, the matrix U obeys the nice equation
\be
UY\llap/=v\llap/U,\;\;v_i=\Delta^{\beta n/2-1}
{(\Pi^{-1})_i}^AY_A
\ee
which probably has the most general solution very similar to the one found in 
\cite{nvv2} for the $AdS_7\times S_4$ case, but we did not investigate that 
further. 

We have also found the full ansatz for the vielbein in the $AdS_5\times 
S_5$ case, as well as the U equation. While these probably have less 
practical applications, they will be useful if one wants to check the 
consistency of the $AdS_5\times S_5$ KK reduction, which is the one thing 
notably missing in the bussiness of consistent truncations
(it's the case most used for the AdS-CFT correspondence). 
We note that if one 
does not restrict to the $Sl(6)/SO(6)$ scalar coset, the equation for the
matrix U does not take a nice form, but instead is found in (\ref{5dueq}).\\
{\bf Note added} After the first version of the paper was posted, we became 
aware of the paper \cite{w}, where the form of the 5d metric (\ref{me}) 
was conjectured.\\
{\bf Acknowledgements} We are grateful to Peter van Nieuwenhuizen for 
discussions and for the the collaboration to an earlier paper \cite{nvv2}, 
from where
most of the work reported here originated. We are also grateful to 
Chris Pope for a discussion of his work, which led us to the analysis in
section 2, and to Radu Roiban for discussions of our results.


\appendix1
\subsection{Killing spinors on spheres}
A Killing spinor $\eta(x)$ on a sphere $S_n$ parametrized by the
coordinates $x^\mu$ is defined by:
\be
D_\mu \eta(x)= c m \gamma_\mu \eta(x) \ee where we introduced a
dimensionful constant $m$, and $c$ is a another dimensionles
constant which will be fixed soon. Spheres are Einstein spaces,
and moreover, are maximally symmetric
\be
{R_{\mu\nu}}^{mn}= m^2(e_\mu^m(x) \;e_\nu^n(x)-e_\mu^n(x)\;
e_\nu^m(x)) \ee Then, the integrability condition reads
\be
[D_\mu , D_\nu]\eta=\frac{1}{4} {R_{\mu\nu}}^{mn} \gamma_{mn}
\eta= \frac{m^2}{2} \gamma_{\mu\nu}\eta=-2 c^2 m^2
\gamma_{\mu\nu}\eta \ee and we derive that $c=\pm i/2$\footnote{In
general, using the integrability condition we derive that Killing
spinors exist only on Einstein spaces, namely $R_{\mu\nu}=4 c^2
(n-1) g_{\mu\nu}$.}. Corresponding to the sign of $c$ we define
two sets of Killing spinors, $D_\mu \eta_\pm =\pm1/2 m \gamma_\mu
\eta_\pm$. Spheres are the only manifold where we can define both
sets $\eta_\pm$. An explicit representation can be given in terms
of stereographic cordinates (see \cite{trieste}): 
\bea
e_\mu^m(z)&=&\frac{\delta^m_\mu}{1+z^2}\;\;\;\;\;\;
\omega_\mu^{mn}(z)=\frac{-2\delta_\mu^m z^n +2\delta_\mu^n
z^m}{1+z^2}\\ \eta_\pm(z)&=&\frac{1\pm i \gamma_m \delta_\mu^m
z^\mu}{\sqrt{1+z^2}} \eta_\pm(0) \label{kspin} 
\eea 
We define further
\be
\bar\eta\equiv \eta^T C \ee where $C$ is the charge conjugation
matrix: $\gamma_\mu^T=\lambda C\gamma_\mu C^{-1}$ and
$C^T=\epsilon C$.
\\
$
\begin{array}{|l|c|c|c|c|c|}
\hline S_n&\epsilon & \lambda &\bar\eta^I \eta^J & gauge
\;\;group&number\;\; of \;\;{\eta_+} 's\\ \hline
n=2&\pm&\pm&\epsilon^{IJ}&SO(3)\simeq SU(2)&2\\
n=3&-&-&\epsilon^{IJ}&SO(4)\simeq SU(2)\times SU(2)&4\\ 
n=4&-&\pm&\Omega^{IJ} &SO(5)\simeq USp(4)&4\\ 
n=5&-&+&\Omega^{IJ}&SO(6)\simeq SU(4)&8\\
n=6&\pm&\pm&\delta^{IJ}&SO(7)&8\\
n=7&+&-&\delta^{IJ}&SO(8)&8\\ n=8&+&\pm&\delta^{IJ}&SO(9)&8\\
n=9&+&+&\delta^{IJ}&SO(10)&16\\ \hline
\end{array}
$
\\
The labeling index $I$ is in the spinor representation of the
gauge group. 

For example, if $\gamma_\mu^T = - C\gamma_\mu
C^{-1}$, then $\bar\eta_{\pm}^I(z)\eta_\pm^J(z)=
\bar\eta_\pm^J(0)\eta_\pm^I(0)$ and the Killing spinors are
normalized to either $\Omega^{IJ}$ or $\delta^{IJ}$ (if $C_{n+1}=C_{(-)}$, 
or $C_{(+)}$ respectively) depending on
the sign of $\epsilon$: $\bar\eta^I
\eta^J=(\bar\eta^I\eta^J)^T=\epsilon\bar\eta^I\eta^J$. In this case, we
can express the Cartesian coordinates $Y^A$ parametrizing the
sphere (with $Y^A$ in the vector representation of the gauge
group) as:
\be
Y^A=\Gamma^A_{IJ} \bar\eta^I_+ \eta^J_-\label{cart} \ee where in
even dimensions $\Gamma^A=\{i\gamma_m\gamma_5, \gamma_5\}$, and in
odd dimensions we use the $\Gamma^A$ in the chiral representation
of $SO(n+1)$. Substituting (\ref{kspin}) into (\ref{cart}) we can
check that with an appropriate choice of $\eta_+(0)$ and
$\eta_-(0)$ we indeed find $Y^A Y^A=1$.

A similar analysis can be done for the case $\lambda=+1$.

Another useful relation is the completeness relation of the Clifford
algebra matrices:
\be
{\gamma^\mu}_{\alpha\beta} {\gamma^\mu}_{\gamma\delta}=
2^{[n/2]-1}\lambda (C_{\alpha\gamma}C_{\beta\delta}+\lambda\epsilon
C_{\beta\gamma}C_{\alpha\delta})+A(\lambda+1)\epsilon C_{\alpha\beta}
C_{\gamma\delta}\label{compln}
\ee
where $2A =\epsilon n -2^{[n/2]-1}\lambda (1+2^{[n/2]}\lambda\epsilon)$.


\begin{thebibliography}{99}
\bibitem{dwn84} B. de Wit, H. Nicolai, Nucl.Phys. B 243 (1984) 91
\bibitem{dwnw} B. de Wit, H. Nicolai and N. P. Warner, Nucl.Phys. B 255
(1985) 29
\bibitem{dwn86} B. de Wit, H. Nicolai, Nucl.Phys. B 274 (1986) 363
\bibitem{dwn87} B. de Wit, H. Nicolai, Nucl.Phys. B 281 (1987) 211
\bibitem{nvv} H.Nastase, D.Vaman and P.van Nieuwenhuizen, hep-th/9905075
\bibitem{nvv2} H.Nastase, D.Vaman and P.van Nieuwenhuizen, hep-th/9911238
\bibitem{mald} J. Maldacena, Adv. Theor. Math. Phys. 2 (1998) 231,
 hep-th/9711200
\bibitem{gkp} S. Gubser, I.R. Klebanov, A.M. Polyakov, Phys. Lett. B428 (1998)
 105, hep-th/9802109
\bibitem{witten} E. Witten, Adv. Theor. Math. Phys. 2 (1998) 253,
hep-th/9802150 
\bibitem{3point}W. Muck, K.S. Viswanathan, Phys. Rev. D58 (1998), 041901
 D.Z. Freedman, S.D. Mathur, A. Matusis, L. Rastelli,
Nucl. Phys. B546 (1999) 96, H. Liu, A.A. Tseytlin, Nucl. Phys. B533 (1998) 88,
 G. Chalmers, H. Nastase, K. Schalm, R. Siebelink,
Nucl.Phys. B540 (1999) 247, G. Arutyunov, S. Frolov, Phys. Rev D 60 (1999)
026004
\bibitem{4point} H. Liu, A.A. Tseytlin,  Phys. Rev D (1999) 086002, 
D.Z. Freedman, S.D. Mathur, A. Matusis, L. Rastelli, Phys. Lett. B 452 (1999) 
61, G. Chalmers, K. Schalm, Nucl. Phys. B554 (1999) 215 and hep-th/9901144,
E. D'Hoker, D.Z. Freedman, Nucl. Phys. B 544 (1999) 612 and Nucl.Phys. B 550
(1999) 261
E. D'Hoker, D. Freedman, S. Mathur, A. Matusis, L. Rastelli, hep-th/9903196
\bibitem{asmo} O. Aharony, S. S. Gubser, J. Maldacena and H. Ooguri,
hep-th/9905111
\bibitem{cdh} M. Cvetic, M.J. Duff, P. Hoxha, J.T. Liu, H. L\"{u}, J.X. Lu,
R. Martinez-Acosta, C.N. Pope, H. Sati and T.A. Tran, hep-th/9903214
\bibitem{cllp} M. Cvetic, J.T. Liu, H. L\"{u} and C.N. Pope, hep-th/9905096
\bibitem{lp} H.L\"u and C.N.Pope, hep-th/9906168
\bibitem{cglp} M.Cvetic, S.S.Gubser, H.L\"u and C.N.Pope, hep-th/9909121
\bibitem{cjlp} E. Cremmer, B. Julia, H. L\"{u} and C. N. Pope, hep-th/9909099
\bibitem{clp} M Cvetic, H. L\"u and C.N. Pope, hep-th/9910252
\bibitem{volk} M. Volkov, hep-th/9910116
\bibitem{chavo} A.H. Chamseddine and M.S. Volkov, Phys. Rev. D 57 (1998),
6242
\bibitem{lpt} H. L\"{u} and C. N. Pope and T.A. Tran, hep-th/9909203
\bibitem{cs} A.H.Chamseddine and W.A.Sabra, hep-th/9911180
\bibitem{rs1} L. Randall, R. Sundrum, hep-th/9905221
\bibitem{rs2} L. Randall, R. Sundrum, hep-th/9906064
\bibitem{bbs} I. Bakas, A. Brandhuber, K. Sfetsos, hep-th/9912132
\bibitem{wz} M. Wijnholt, S. Zhukov, hep-th/9912002
\bibitem{kl} R. Kallosh, A. Linde, hep-th/0001071
\bibitem{bc} K. Behrndt, M. Cvetic, hep-th/9909058, hep-th/0001159
\bibitem{anp} M.A. Awada, B.E.W. Nilsson, C.N. Pope, Phys. Rev. D 29 (1984)
334
\bibitem{nilsson} B.E.W. Nilsson, Phys. Lett. 155B (1985) 54
\bibitem{ppn} M. Pernici, K. Pilch and P. van Nieuwenhuizen, Phys.Lett. 143 B
(1984) 103
\bibitem{nt} P.K.Townsend, K.Pilch and P.van Nieuwenhuizen, Phys. Lett. B 136,
38 (1984)
\bibitem{tn83} P.K.Townsend and P.van Nieuwenhuizen, Phys.Lett. 125, 41 (1983) 
\bibitem{cj} E.Cremmer and B.Julia, Nucl.Phys. B 141 (1979) 141
\bibitem{clps} E. Cremmer, H. L\"u, C.N. Pope and K.S. Stelle,
hep-th/9707207
\bibitem{trieste} P.van Nieuwenhuizen, Proceedings of the Trieste Spring School
1984, p.239
\bibitem{w} A. Khavaev, K. Pilch, N. Warner, hep-th/9812035
\end{thebibliography}
\end{document}